\documentclass[fleqn,usenatbib]{mnras} 
\usepackage{mathptmx}
\usepackage[T1]{fontenc}
\usepackage{ae,aecompl}
\usepackage{times}
\usepackage{amssymb}
\usepackage{amsmath}
\usepackage{upgreek}
\usepackage{graphicx}
\usepackage{lscape}
\usepackage{float}
\usepackage{longtable} 
\usepackage[usenames, dvipsnames]{color}

\usepackage{enumitem} 
\setlist{itemsep=4pt} 

\title[Ly$\alpha$ LF from MUSE] 
{MUSE Deep-Fields: The Ly$\alpha$ Luminosity Function  {\mbox{in the Hubble
  Deep Field South}} at $ 2.91 < z
  < 6.64$.}
\author[Alyssa B. Drake et al.]{Alyssa B. Drake$^{1}$, Bruno
  Guiderdoni$^{1}$, J\'{e}r\'{e}my
  Blaizot$^{1}$, Lutz
  Wisotzki$^{2}$, \newauthor Edmund Christian Herenz$^{2}$, Thibault
  Garel$^{1}$, Johan Richard$^{1}$, Roland Bacon$^{1}$, \newauthor David Bina$^{5}$,  Sebastiano Cantalupo$^{3}$, Thierry
  Contini$^{5,6}$, Mark den Brok$^{3}$, \newauthor Takuya Hashimoto$^{1}$,
  Raffaella Anna Marino$^{3}$, Roser Pell\'{o}$^{5}$, Joop
  Schaye$^{4}$, \newauthor Kasper B. Schmidt$^{2}$ 
\\ $^{1}$Univ Lyon, Univ Lyon1, Ens de Lyon, CNRS, Centre de Recherche
Astrophysique de Lyon UMR5574, F-69230, Saint-Genis-Laval, France
\\$^{2}$ Leibniz-Institut f\"{u}r Astrophysik Potsdam (AIP), An der
Sternwarte 16, 14482 Potsdam, Germany
\\$^{3}$ ETH Z\"{u}rich, Institute for Astronomy, Department of
Physics, Wolfgang-Pauli-Strasse 27, 8093 Z\"{u}rich, Switzerland
\\$^{4}$ Leiden Observatory, P.O. Box 9513, NL-2300 RA Leiden, The Netherlands
\\$^{5}$ IRAP, Institut de Recherche en
Astrophysique et Plan\'{e}tologie, CNRS, 14 avenue \'{E}douard Belin,
31400 Toulouse, France, 
\\$^{6}$ Universit\'{e} de Toulouse, UPS-OMP, 31400 Toulouse, France}

\begin{document}               

\pagerange{\pageref{firstpage}--\pageref{lastpage}} \pubyear{2015}

\maketitle

\label{firstpage}

\begin{abstract}\\
We present the first estimate of the Ly$\alpha$ luminosity function
using blind spectroscopy from the {\emph{Multi Unit Spectroscopic
    Explorer}}, MUSE, in the {\emph{Hubble Deep Field South}}. Using automatic
source-detection software, we assemble a homogeneously-detected sample
of 59 Ly$\alpha$ emitters covering a flux range of $-18.0 <
{\rm{log_{10}}}\; (F) < -16.3\; ({\rm{erg\; s}}^{-1} {\rm{cm}}^{-2}$),
corresponding to
luminosities of $41.4 < {\rm{log_{10}}}\; (L) < 42.8\;
({\rm{erg\; s}}^{-1})$. As recent
studies have shown, Ly$\alpha$ fluxes can be underestimated by a
factor of two or more via traditional methods, and so we undertake a careful
assessment of each object's Ly$\alpha$ flux using a
curve-of-growth analysis to account for extended emission. We describe
our self-consistent method for determining the completeness of the
sample, and present an estimate of the global Ly$\alpha$ luminosity
function between redshifts $2.91 < z < 6.64$ using the $1/V_{max}$
estimator. We find the luminosity function is higher than many
number densities reported in the literature by a factor of $2-3$, although our
result is consistent at the $1\sigma$ level with most of these
studies. Our observed luminosity function is also in good agreement
with predictions from semi-analytic models, and shows no
evidence for strong evolution between the high- and low-redshift halves of the data. We
demonstrate that one's approach to Ly$\alpha$ flux estimation does alter
the observed luminosity function, and caution that accurate flux
assessments will be crucial in measurements of the faint end slope. This is a pilot study for the
Ly$\alpha$ luminosity function in the MUSE deep-fields, to be built on
with data from the {\mbox{{\emph{Hubble Ultra Deep
Field}}}} which will increase the size of our sample by almost a
factor of $10$. 

\end{abstract}

\begin{keywords}
cosmology:observations - surveys - galaxies:evolution -
galaxies:formation - galaxies:high-redshift - galaxies:luminosity functions.
\end{keywords}

\section{Introduction}
\label{intro}
The Ly$\alpha$ emission line is one of the most powerful probes of the
early Universe, giving us insight into the very early stages of
galaxy formation. Galaxies detected via their Ly$\alpha$-emission {\mbox{(LAEs;
\citealt{Cowie&Hu1998})}} offer us a means
to study high-redshift star-forming galaxies, even with continuum
magnitudes too faint to be observed using current technology. These
low-mass objects form the building blocks of {\mbox{$L$$^*$ galaxies}} in the
local Universe ({\mbox{\citealt{Dayal2012}}}, \citealt{Garel2016}), meanwhile theoretical models
suggest they may also play a significant role in driving cosmic
reionisation e.g. \cite{Gronke2015a}, \cite{Dijkstra2016}, \cite{Santos2016}.

Although Ly$\alpha$ physics
is complex (e.g. \citealt{Verhamme2006}, \citealt{Gronke2015b}), we
can begin to understand the physical processes underway at these epochs by measuring the luminosity function of LAEs -- a fundamental statistic of the
population. The
luminosity function tells us about the relative
abundance of different luminosity objects in the overall distribution
(e.g. see \citealt{Johnston2011}),
and ultimately for deep enough samples, measurement of the faint-end slope will tell us if LAEs
are numerous enough to be the primary sources of reionisation (see
\citealt{Dressler2015} for a discussion).

The largest samples of LAEs to date come from narrow-band selection, whereby a
narrow filter (typically $< 100$ \AA), is used in combination
with broad-band photometry to detect emission-line galaxies in distinct redshift ``slices''. This
approach is very efficient, having led to samples of thousands
of H$\alpha$, H$\beta$, [O{\sc{iii}}] and [O{\sc{ii}}] emitters out to
$z\approx 2.0$ (\citealt{Sobral2009},
\citealt{Sobral2013}, \citealt{Drake2013}, \citealt{Drake2015}) as
well as LAEs at $z > 3.0$ (\citealt{Rhoads2000}, \citealt{Ouchi2003},
\citealt{Hu2004}, \citealt{Ouchi2008}, \citealt{Yamada2012}, \citealt{Matthee2015},
\citealt{Konno2015}, \citealt{Santos2016}). These relatively shallow surveys have provided increasingly robust estimates of the
Ly$\alpha$ luminosity function down to
luminosities of log$_{10}$ $L\approx 42.0\; {\rm{erg\, s}}^{-1}$, in the redshift
interval $\approx 2.0 < z < 7.0$. Typically these studies estimate values of the
characteristic number density and luminosity of the sample, although
the faint-end slope remains unconstrained.

Spectroscopic studies provide an alternative approach, allowing the
identification of LAEs without any need for ancillary
data, but typically surveying far smaller volumes. In addition to
targetted spectroscopy, one can place long-slit spectrographs blindly on sky, but
the results often suffer from
severe slit-losses and a complicated selection function. (See also
survey results from low-resolution slitless spectroscopy \citealt{Kurk2004}, \citealt{Deharveng2008} and IFU studies
\citealt{vanBreukelen2005}, \citealt{Blanc2011}). In recent years, spectroscopic surveys have begun to push
Ly$\alpha$ samples to lower flux limits than ever before, complementing
wide, shallow, studies with very deep integrations. The two deepest such
surveys to date come from \cite{Rauch2008}
and \cite{Cassata2011} reaching 1 dex deeper than their
narrow-band counterparts. \cite{Rauch2008} used a 92
hour long-slit exposure with the ESO VLT FORS2 instrument, detecting single-line emitters of just a few $\times 10^{-18} {\rm{erg\,s}}^{-1} {\rm{cm}}^{-2}$
corresponding to Ly$\alpha$ luminosities of $\approx 8 \times 10^{40} {\rm{erg\,s^{-1}}}$ for
LAEs in the range $2.67 < z < 3.75$. The
authors note however that their luminosities could be underestimated
by factors of $2 –- 5$ due to slit
losses, and the identification of many of their single-line emitters
is somewhat uncertain. Another notable study came from the VIMOS-VLT
Deep Survey (VVDS; \citealt{Cassata2011})
finding $217$ LAEs with secure spectroscopic redshifts between $2.00 <
z < 6.62$, and fluxes reaching as low as $F = 1.5 \times 10^{−18}
{\rm{erg\, s}}^{-1} {\rm{cm}}^{-2}$. The detections came from a
combination of targetted and serendipitous spectroscopy however, and again resulted in
a complex selection function and slit losses. Nevertheless, the number of emitters in their
sample allowed the authors to split the data into three
redshift bins, to look for any sign of evolution in the observed
luminosity function. They ultimately found no evidence in support of evolution,
consistent with the previous results of \cite{vanBreukelen2005},
\cite{Shimasaku2006}, and \cite{Ouchi2008}. Finally, at the highest redshifts, the first robust constraints on the faint end of the Ly$\alpha$ luminosity
function came from \cite{Dressler2015}. They found a very steep value of the faint-end
slope at $z = 5.7$, using targets selected via ``blind long-slit
spectroscopy'', further reinforcing the significance of intrinsically faint LAEs
in the early Universe (see also \citealt{Dressler2011} and \citealt{Henry2012}).

The low--luminosity LAE population is now at the forefront of
research, meaning that the accurate recovery of total LAE fluxes is of
high priority for upcoming work. Indeed, some studies have already suggested that all LAEs
exhibit extended, low-surface-brightness Ly$\alpha$ emission coming
from the surrounding circum-galactic medium. The detection
of this emission is difficult, and requires very
sensitive measurements indeed. \cite{Momose2014} built on the work of
\cite{Matsuda2012} by stacking LAE detections in $5$ redshift
slices between $\approx 2.2 < z < 6.6$ resulting in the detection of
extended Ly$\alpha$ emission around normal star-forming galaxies
across this entire epoch. They found typical exponential scale
lengths of $\approx5 -– 10$ kpc, but the emission was not detectable
around any individual galaxy (see also \citealt{Yuma2013} for
indivisual detections of metal-line
blobs at lower z).

The Multi Unit Spectroscopic Explorer (MUSE; \citealt{Bacon2010}) on the Very Large
Telescope (VLT) allows us to carry out blind spectroscopic selection of LAEs between
redshifts $\approx 3.0 < z < 6.5$ with a homogeneous selection function. The efficiency of this approach to detect line emission allows us to use MUSE as a detection
machine for the kind of star-forming galaxies we wish to
trace, also enabling an accurate assessment of total Ly$\alpha$ fluxes. 
\cite{Bacon2015}, hereafter B15, presented a
blind-spectroscopic  analysis of the {\mbox{{\emph{Hubble Deep Field South}} (HDFS)}},
and the resultant catalogue showcased the detection power of
MUSE. Indeed, B15 presented several galaxies detected via their line
emission alone, that were otherwise undetectable in the deep broad-band
HST imaging  (I$_{814} > 29$ mag AB). Additionally, MUSE is able to overcome the
effects of slit-loss that have so far hampered Ly$\alpha$ flux estimates
from long-slit specroscopy, allowing us to perform a careful evaluation of the total Ly$\alpha$ flux from each
    galaxy. For instance, \cite{Wisotzki2016} used a curve-of-growth
    analysis on $26$ isolated halos in the B15 catalogue, and presented the first ever detections of extended
Ly$\alpha$ emission around individual, high-redshift, star-forming
galaxies. The objects presented were in the flux range $ 4.5
\times 10^{-18} {\rm{erg\, s}}^{-1} {\rm{cm}}^{-2}$ up to  $3
\times 10^{-17} {\rm{erg\, s}}^{-1} {\rm{cm}}^{-2}$ across the redshift interval  $2.96 < z < 5.71$, and
halos were detected around $21$ of these objects. The omission of this low surface brightness
contribution to the total Ly$\alpha$ flux has potentially led to a
systematic underestimation of Ly$\alpha$ fluxes in the literature, and
lends support to the importance of a re-assessment of the
Ly$\alpha$ luminosity function.

In this paper we present a pilot study for the LAE luminosity
function using blind spectroscopy in the 1 square arcminute HDFS field. We use automatic detection software to present a
homogeneously selected sample of 59 LAEs and estimate Ly$\alpha$
fluxes via a curve-of-growth analysis to account for extended Ly$\alpha$ emission. We have developed
and implemented a self-consistent method
to determine the completeness of our sample, allowing us to compute a global Ly$\alpha$ luminosity function using the $1/V_{max}$ estimator.

The outline of this paper is as follows. In Section
\ref{sect:data} we present our observations from MUSE and outline our
method of catalogue construction and sample selection. In Section
\ref{sect:flux} we describe our approach to estimating
the Ly$\alpha$ flux, and in Section \ref{sect:compl} we present and
discuss our completeness estimates for the sample. In
Section \ref{sect:results} we present our estimation of the
LAE luminosity function between $2.91 < z < 6.64$, and discuss our results in the context of
observational literature as well as in comparison to the semi-analytic model of \cite{Garel2015}. In Section
\ref{sect:disc} we examine the effect of using different flux
estimates for LAEs and look for evolution over the redshift range of
our observed luminosity function. Finally, we summarise our results in {\mbox{Section
\ref{sect:concl}}}.

\begin{figure*}
\begin{center}
 \includegraphics[width=0.48\textwidth]{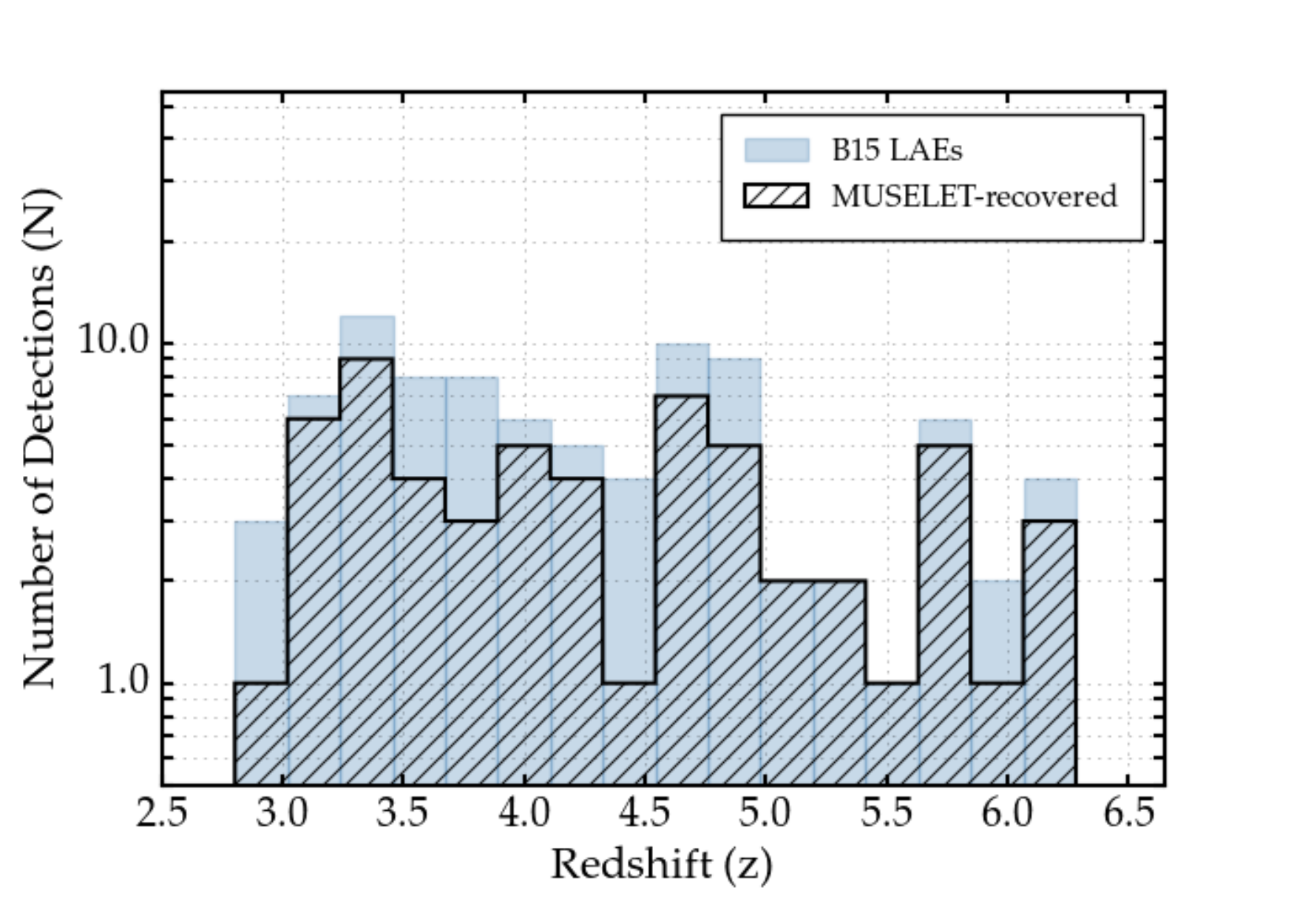} 
 \includegraphics[width=0.48\textwidth]{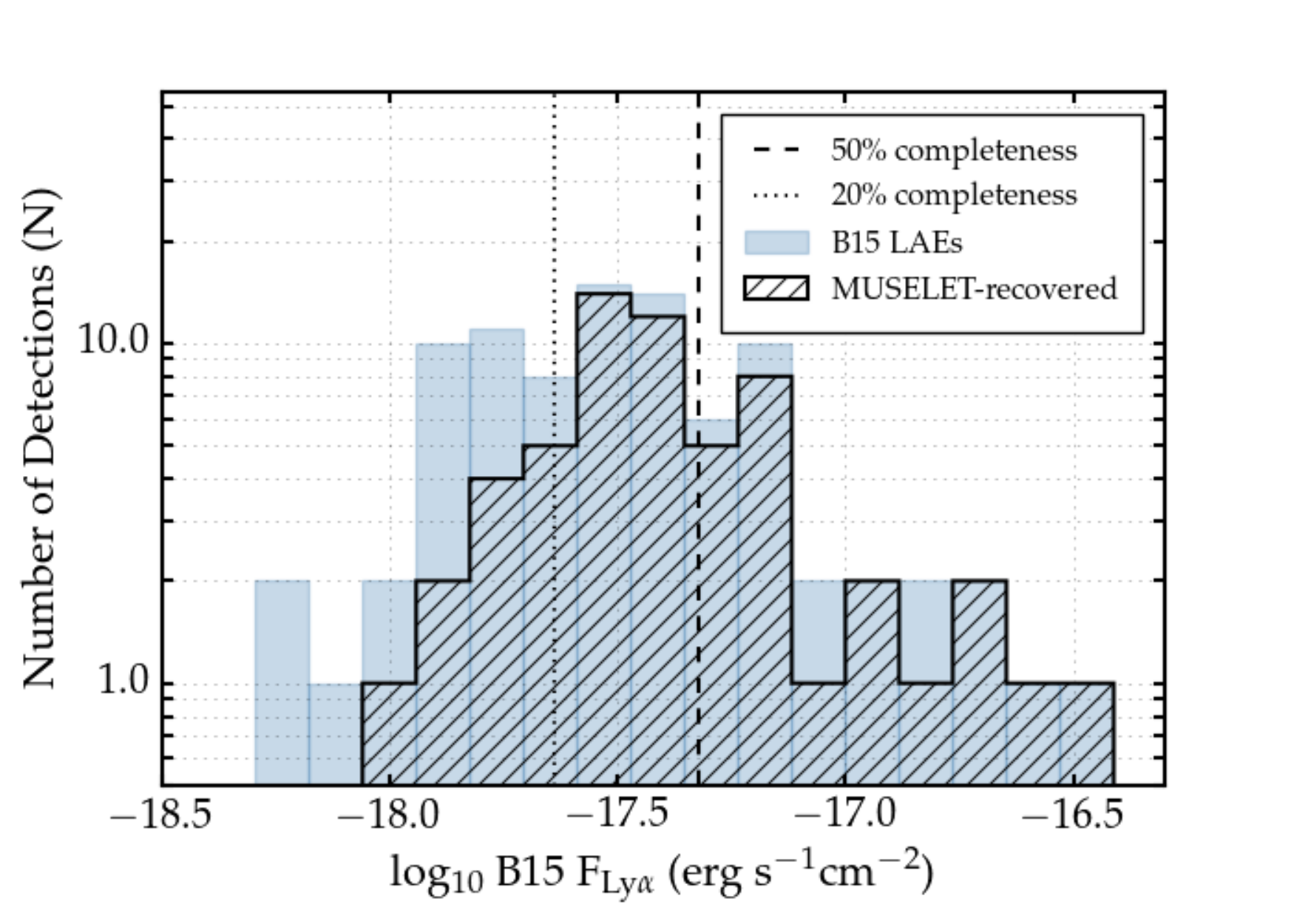} 
 \caption{A comparision between numbers of LAEs presented in Bacon
   et al. (2015) and detections recovered using the detection software
   {\sc{muselet}}. In the left-hand panel we show the redshift distribution
   of our detections overlaid on the redshift distribution of the B15
   LAEs. This demonstrates an even recovery rate across the entire redshift
   range i.e. no redshift bias in our method of detection. In the
   right-hand panel we use the published flux estimates of B15 to show the distribution of fluxes
   recovered by {\sc{muselet}} vs the distribution for B15 LAEs. We
   successfully recover the majority of bright LAEs before
   incompleteness becomes more apparent below {\mbox{log$_{10}$ $F$
       Ly$\alpha$ (B15) $=
-17.32$}}. Bright LAEs which are not recovered by {\sc{muselet}} lie
in the small parts of the cube with fewer than 50 percent of the final
exposure time. The average sample completeness is overlaid (dashed and
   dotted lines) and its derivation is described in \mbox{Section \ref{sect:compl}}.}
  \label{fig:no_counts}
\end{center}
\end{figure*}

The total co-moving volume between {\mbox{$2.91 <
  z < 6.64$}} equates to {\mbox{$10351.6$ Mpc$^3$}}. As parts of the cube are excluded from the
search however (see Section \ref{subsect:det}) the total co-moving
survey volume is reduced to {\mbox{$10144.57$
Mpc$^3$}}.
Throughout this paper we assume a $\Lambda$CDM cosmology, \mbox{$H_0 =
  70.0$} kms$^{-1}$ Mpc$^{-1}$,
$\Omega_{m} =0.3$, $\Omega_\Lambda = 0.7$.

\section{Data and Sample Selection}
\label{sect:data}

\subsection{Observations and Data Reduction}
\label{obs}
During the final MUSE commissioning run in
July 2014, we performed a deep integration on the HDFS for a total of 27 hours, using the standard wavelength
range 4750-9300 \AA. Seeing was good for most nights ranging between
0.5-0.9 arcsec. The full details of these observations are
given in B15.

We use a new reduction of the cube optimised for the
detection of faint emission-line objects (v1.4; Cantalupo, in prep). The reduction uses the CubExtractor
package and tools to minimise residuals around bright sky lines. For a more detailed description of the flat-fielding and sky-subtraction procedures with the CubExtractor package see, e.g.\,\cite{Borisova2016}. A detailed comparison between this improved reduction for the HDFS field with respect to previous versions will be presented in Cantalupo et al., in prep.

\subsection{Catalogue Construction}
\label{cat}

When assessing the luminosity function, it
is of fundamental importance to understand the selection
function of the galaxies which make up the sample. This means that the catalogue of LAEs
must be constructed homogeneously, and in a 
way which allows us to assess the completeness of the sample in a consistent manner.

We therefore choose to implement a single
method of source detection allowing us to apply homogeneous selection
criteria across the field, and to apply these same criteria in our
fake source recovery experiment (see Section \ref{sect:compl}). We
highlight here that any automated catalogue construction will require
some trade-off to be made between the depth of the catalogue and the
false detections which are included. In this work, we choose a
conservative setup of our detection software to minimise false
detections, resulting in a very robust selection of objects.

Finally, one needs to verify the nature of each source as an LAE, and for this
we rely on the deeper catalogue presented in B15 (details below). This means that by construction, our catalogue will
always form a subsample of B15. While the B15 catalogue is deep and meticulously constructed,
the objects were detected through a variety of means, and
the heterogenetity of the sample results in an irregular selection
function which would be impossible to reproduce. For this reason the
B15 catalogue is unsuitable for the
construction of a luminosity function.

\subsubsection{Source Detection}
\label{subsect:det}
Our chosen software, ``{\sc muselet}''
(J. Richard), has been optimised for the detection of line
emission, and has been extensively tested on both blank and cluster
fields. {\sc muselet} makes extensive use of the {\sc SExtractor} package
\citep{BertinArnouts1996} to perform a systematic search through the data
cube for emission-line objects. The input data cube is manipulated to create a
 continuum-subtracted narrow-band image at each wavelength plane.
Each narrow band image is based on a line-weighted average of 5
wavelength planes in the cube ($6.25$\AA\; total width), and the continuum is
estimated from $2$ spectral medians of $\approx 25$\AA\; on each the blue
and the red side of the narrow band region. {\sc SExtractor} is run
on each of these images as they are created\footnote{{\sc SExtractor}
  parameters are set to {\sc detect minarea} $= 3.0$, and
{\sc detect thresh} $= 2.5$. These are the minimum number of pixels above the
threshold and the sigma of the detection respectively.}, using the exposure map
cube as a weight map, and rejecting all detections in areas of the cube with fewer than $50$
percent of the total number of exposures. This reduces the volume
probed to $0.98$ of the full cube, and is taken into
account in the construction of the luminosity function. Once the entire cube has
been processed, {\sc muselet} merges all of the {\sc SExtractor}
catalogues, and records a detection at the wavelength of the peak of
the line. This results in a ``raw'' catalogue of emission
lines.

\subsubsection{Candidate LAE Selection}

{\sc muselet} includes the option to interpret this raw catalogue of detections
as individual objects. Using an input list of rest-frame emission
line wavelengths and flux ratios, we can combine lines co-incident
on-sky, and estimate a best redshift for each object showing multiple
emission peaks. Emission lines are merged spatially into the same source
based on the ``radius'' parameter (here radius $= 4$ pixels or $0.8$ arcseconds), and the object
must be detected in two consecutive narrowband images in the cube to
register as a real source.

 Thanks to the wavelength coverage and sensitivity
    of MUSE, we anticipate the detection of multiple lines for 
    galaxies exhibiting any of the major emission lines associated
    with star formation. Only
those sources exhibiting a single emission line are flagged as ``Ly$\alpha$/[O{\sc{ii}}]''
emitters for validation. This equates to $144$ single-line sources.

\subsubsection{LAE Verification}
We now have a robustly detected catalogue of single line emitters, and we rely on the detailed work presented in
B15 to give us a means to distinguish between
Ly$\alpha$ and [O{\sc{ii}}] emitters. Of the $144$ single line
emitters detected with {\sc muselet}, $59$ are
identified as LAEs through careful matching to B15. To qualify as a
match to the B15 catalogue, the positions on sky must lie within a $1.0$
arcsecond radius of one another and within $6.25$ \AA\, in
wavelength. 
The B15 catalogue was constructed taking full advantage of the deep
HST imaging across the field, initially extracting spectra at the
positions of objects presented in the HST catalogue of {\mbox{\cite{Casertano2000}}}. In a
complementary approach, several pieces of
detection software\footnote{SExtractor; \cite{BertinArnouts1996}, LSDcat; Herenz et
al. (in prep)} were used to search for pure
emission-line objects as liberally as possible, as well as several searches conducted by
eye. Via each of these methods, all detections were scrutinised by at least two authors of B15, comparing
spectral extractions, narrow-band images and HST data before the
object was validated.

\subsubsection{Final Catalogue}

In the left-hand panel of Figure \ref{fig:no_counts} we show the
redshift distribution of the $89$ B15 LAEs with secure identifications
(Q $>= 2$)\footnote{Confidence levels in B15 range between Q=0 (no
  secure redshift) and Q=3 (redshift secure and based on mulitple
  features). Q=2 refers to a single-line redshift with a high
  signal to noise (i.e. to distinguish between the Ly$\alpha$ and
  [O{\sc{ii}}] line profiles).} according to the assigned confidence
level in B15. Overlaid is the distribution
of the {\sc muselet}-selected LAE sample which match to existing objects 
in B15. We find recovery is evenly distributed across the entire
redshift range in the deeper B15 catalogue, indicating no redshift bias
in our object detection. In the right-hand panel
of Figure \ref{fig:no_counts}, we show the distribution of fluxes
reported in B15 for the same two samples. Ly$\alpha$ flux values in B15 come from
{\sc platefit} \citep{Tremonti2004} $1$D spectral extraction
estimates, using a Gaussian profile fit to the Ly$\alpha$
line. We note that this is not the optimal procedure to estimate
Ly$\alpha$ flux, and we do not use these values in the determination
of the luminosity function or the remainder of this paper - see {\mbox{Section \ref{sect:flux}}}
for a discussion of the factors affecting flux estimation and a
description of our improved approach. 

We recover almost all LAEs with a B15 flux greater than the average
$50$ percent sample completeness limit at {\mbox{log$_{10}$ B15 $F$ Ly$\alpha =
-17.32$}} (see Section \ref{sect:compl}). We miss only those that lie in parts
of the cube with fewer than $50$ percent of the total exposure time
which are rejected by {\sc{muselet}}, but seen by eye or alternative software in
B15. We detect $24$ of the $26$ bright isolated LAEs presented in
\cite{Wisotzki2016} which were drawn from the B15 sample. On visual
inspection these two objects, although bright, are found in the very
small parts of the cube with less than 50 percent of the total
exposure time, and therefore are not recovered with the chosen
{\mbox{{\sc{muselet}}}} setup. 

For the remainder of the analysis, we make the assumption that any
{\sc muselet} single line detections which are
not verified as LAEs by the extensive B15 catalogue are
[O{\sc{ii}}] emitters or spurious detections, and
can be excluded from the analysis.

\section{Fluxes}
\label{sect:flux}

\begin{figure}
\begin{center}
 \includegraphics[width=0.5\textwidth]{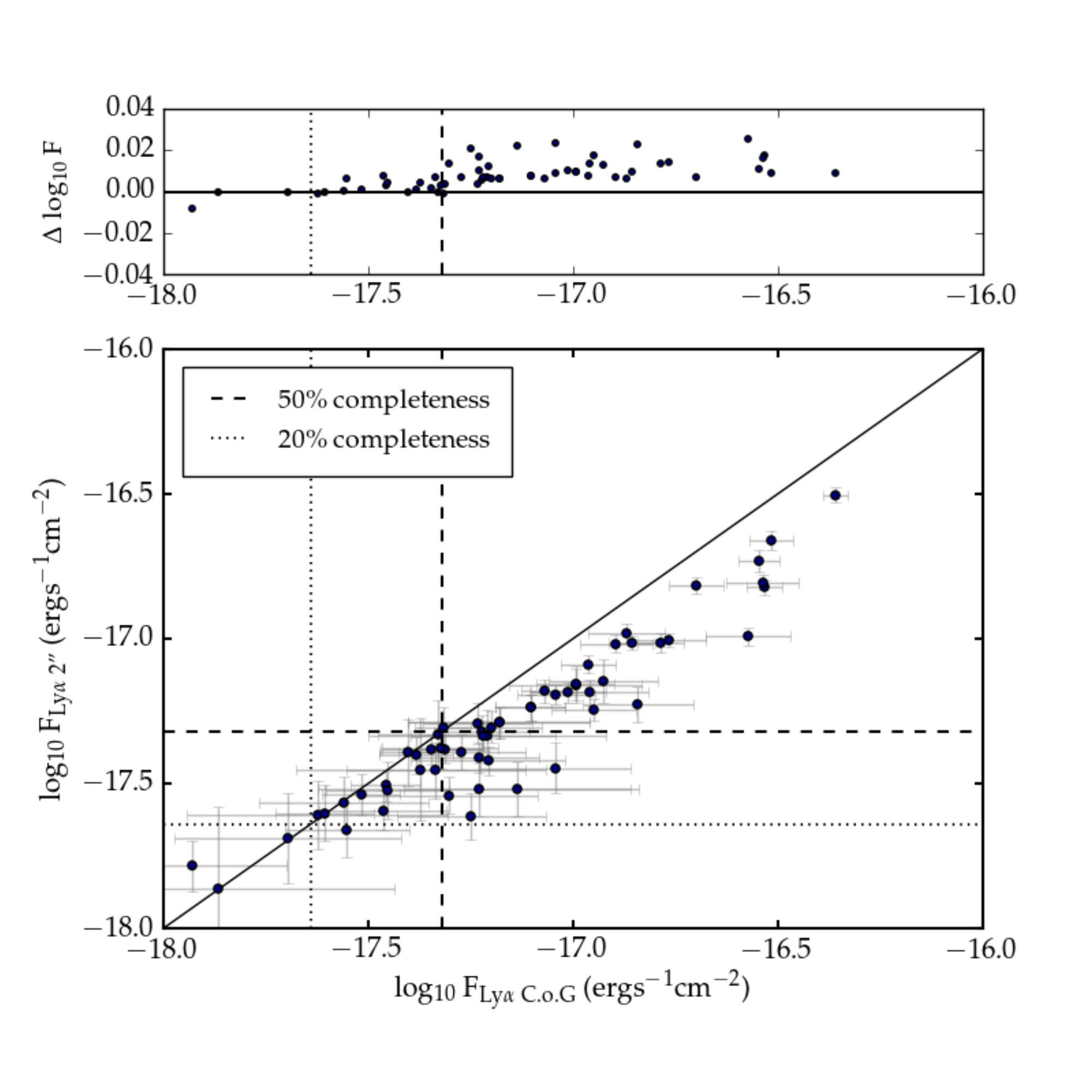} 
 \caption{Comparison of Ly$\alpha$ flux estimates. The upper panel shows {\mbox{$\Delta$ log$_{10}$ F}} as a function of
   $F$$_{{\rm{\;C.o.G}}}$, where {\mbox{$\Delta$ log$_{10}$ F = (log$_{10}$ F$_{2"}$ $-$
   log$_{10}$ F$_{{\rm{C.o.G}}}$) / log$_{10}$ F$_{{\rm{C.o.G}}}$}}. The lower panel shows a direct
   comparison of flux estimates from $F$$_{{\rm{2 \arcsec}}}$ and
   $F$$_{{\rm{\;C.o.G}}}$. Error bars depict the standard
   deviation from pixel statistics on
   each flux measurement. The sample completeness (overplotted dashed and
   dotted lines) is described in
   \mbox{Section \ref{sect:compl}}. While the two estimates agree at fluxes
   lower than {\mbox{log$_{10}$ $F$$_{{\rm{}}} \approx -17.3$}}, brighter than this
   the two measurements deviate increasingly, highlighting the need
   for a careful assessment of total flux when dealing with LAEs.}
  \label{fig:flux}
\end{center}
\end{figure}

The accurate recovery of line fluxes plays an important role in
determining the luminosity function. In addition to the difficulties of flux
measurement from long-slit spectroscopic observations, B15 noted that
even when utilising a data cube, in deep integrations such as these, source
crowding can lead to necessarily small spectral extractions, and hence
the outer parts of extended sources can be unaccounted
for. In the case of the fluxes quoted in B15, the flux underestimate will be
exacerbated in some cases due to the fact that {\sc platefit} was not designed to
deal with LAEs which often exhibit an asymmetric profile. \cite{Wisotzki2016} reported
for instance that Ly$\alpha$ fluxes in B15 from {\sc{platefit}} were sometimes more than a
factor of two too low. 

Our preferred approach is to perform photometry on pseudo narrow-band images constructed by
collapsing several planes of a datacube in the spectral
direction allowing us to treat the outer parts of each source with
greater care. We conduct this analysis in two ways in order to
demonstrate the difference in measured LAE fluxes when working with
different sized apertures to those which have often been used in the
literature.

\subsection{Methods of Ly$\alpha$ Flux Estimation}
For each confirmed LAE, we extract a $1$D spectrum from the cube, using an aperture
defined by the segmentation map from {\mbox{{\sc SExtractor}}}. This spectrum
is used only to gain some measure the FWHM of the line, by fitting a
Gaussian to the profile. Next we extract a ``narrow-band'' image from the cube centred on the
detection wavelength, of width $\Delta\, \lambda = 4 \times
{\rm{FWHM}}$, and a ``continuum image'' on the red side of the line, offset by 50 \AA\,,
and of width $\Delta\, \lambda = 200$ \AA\,. Finally, we subtract the mean
continuum image from the mean narrow-band image to construct a
``Ly$\alpha$ image'' (multiplied by the width of
the narrowband image for correct flux units). We perform all photometry on this final image,
masking objects in close proximity to the LAE, seen in the corresponding
continuum or narrow-band images. 

We consider two different approaches to flux estimation using
aperture photometry on the Ly$\alpha$ images. First we conduct photometry in an aperture of $2$ arcseconds in diameter, and then carry out a
curve-of-growth analysis using the light profile of each object to
judge the appropriate size of aperture to account for extended
emission. To measure the light profile of each object, we centre an
annulus on the object in our masked
Ly$\alpha$ image, before stepping through consecutive annuli of increasing
radii measuring the flux in each ring. The total flux is then determined as the sum of the annuli out to the radius
where the mean flux in an annulus reaches or drops below zero. This is
where the light profile of the object hits the
background of the image. Removing the local background of the
objects made no significant impact on our results. 

\subsection{Comparison of Flux Estimates}

Figure \ref{fig:flux} shows a comparison between the measured
two arcsecond aperture flux, $F_{{\rm{\,2\arcsec}}}$, and the curve of
growth flux, $F_{{\rm{\,C.o.G}}}$. The estimates are in good
agreement below {\mbox{$F$$_{{\rm{}}} \approx -17.3$}} which is also where the
sample reaches an average completeness of 50 percent (see Section
\ref{sect:compl} for details). Upwards of this, $F_{{\rm{\,C.o.G}}}$ starts to deviate more dramatically from
$F_{{\rm{\,2\arcsec}}}$. This means that flux measurements of the
brightest LAEs will differ most according to the approach used,
possibly introducing some bias into measurements of the luminosity
function at different redshifts. We investigate the effect of
different methods of flux estimation on the luminosity function in Section \ref{sect:LF_flux}.

The objects blindly detected by {\sc{muselet}} are summarised in Table
\ref{tab:fluxes} with Ly$\alpha$ flux estimates resulting
from our curve of growth analysis as well as two arcsecond aperture
photometry. Errors on our flux estimates
are given by the standard deviation of each measurement according to
pixel statistics. We also show the published Ly$\alpha$ fluxes from
both B15 and \cite{Wisotzki2016}  where $26$
objects were carefully re-examined.

\section{Sample Completeness}
\label{sect:compl}
\begin{figure*}
\begin{center}
 \includegraphics[width=\textwidth]{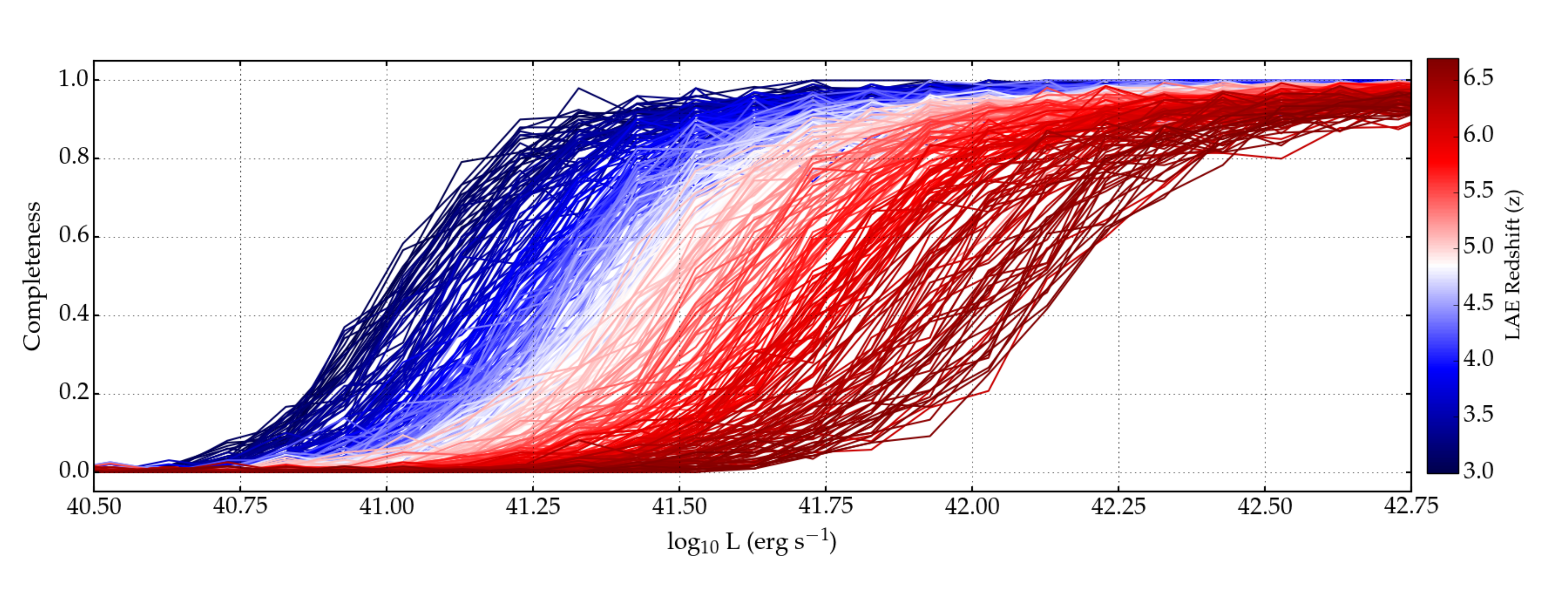} 
 \caption{Completeness as a function of LAE luminosity. We show the
   recovery fraction of LAEs at $40$ different input luminosities,
   colour-coded by redshift in intervals of $\Delta\, z = 0.01$ ($\Delta\,
   \lambda \approx 12$ \AA). At higher redshifts
 LAEs must have a higher luminosity before they can be
 detected. Additionally, the detectabilty of higher redshift LAEs
 increases more slowly with increasing luminosity since night sky emission
 hampers observations towards longer wavelengths.}
  \label{fig:c(z)sky}
\end{center}
\end{figure*}

\subsection{Fake Source Recovery}
To make quantitative measures of the completeness of our LAE sample
from {\sc muselet}, we insert fake
point-source line emitters distributed randomly on-sky into the real
data cube. For each fake line-emitter the properties of the Ly$\alpha$ line profile
(asymmetry and velocity width) are drawn randomly from the measured
profiles of the LAEs presented in B15, and the
objects are required to scatter randomly on-sky with no avoidance of each
other or of real objects. By definition this means that the
completeness estimate will never reach 100 percent as objects can fall
on top of one another, behind real sources, or in the small volume of
the cube where exposure time is less than 50 percent of the total
integration where sources are rejected by {\sc{muselet}}. This allows an
exact imitation of the method via which we construct our catalogue,
and ensures that the two volumes surveyed are identical. 

We work systematically through the data cube inserting $20$ fake LAEs at a time in redshift
bins of $\Delta z = 0.01$ corresponding to wavelength intervals of $\approx 12$ \AA. Each
    point-source LAE is convolved with the MUSE PSF to create a tiny
    cube containing only an LAE spectrum (no continuum emission) and its
    associated shot noise in a variance cube. The mini data- and
    variance cubes are then added directly to the real data and
    variance cubes. Crucially, we make
the assumption here that all input fake LAEs would indeed be
correctly classified by matching to B15.

\subsection{Completeness as a Function of Luminosity}
In Figure \ref{fig:c(z)sky} we show the recovery fraction of LAEs
with {\sc muselet} as a function of log
luminosity. We use $40$ values of log luminosity, and $370$ tiny
redshift bins, showing LAE-redshift in the colour bar. 

In the lowest
redshift bin at $z = 3.00$, we begin to detect objects at {\mbox{log$_{10}$ ($L$$_{{\rm{}}}) \approx 40.65$}},
reaching a $90$ percent recovery rate by {\mbox{log$_{10}$
    ($L$$_{{\rm{}}}) \approx 41.20$}}. By redshift $z = 6.64$ objects are not recovered unless their
luminosity exceeds  {\mbox{log$_{10}$ ($L$$_{{\rm{}}}) = 41.65$}},
reaching $90$ percent completeness by
{\mbox{log$_{10}$ ($L$$_{{\rm{}}}) = 42.60$}}. In addition to
the shift towards brighter luminosities for
each completeness curve with increasing redshift, the gradient of each
curve also gradually decreases with increasing redshift. This behaviour is
due to night sky emission becoming more prominent towards longer
wavelengths, and hampering the detection of even luminous LAEs at higher redshifts. Taking the lowest and
highest redshift bins again, we see that the recovery fraction in the lowest
redshift bin goes from $10$ to $90$
percent across a luminosity interval of $0.40$ dex, whereas at the highest redshifts in
the sample, the same interval in completeness spans a luminosity
range of $0.75$ dex. This reinforces our choice of a very finely
sampled redshift range, as completeness levels
will vary significantly according to the proximity of each LAE's
observed wavelength to sky lines.

In order to approximate the average completeness of the sample in
terms of LAE flux, we can combine these results across all wavelengths. Using the input redshift and luminosity of each fake LAE, we can
determine its flux, and record the information of whether the object
was recovered by {\sc{muselet}} or not. This way we estimate that the sample completeness drops to $50$
percent by {\mbox{log$_{10}$ ($F$$_{{\rm{}}}$) $= -17.32$}} and $20$
percent by {\mbox{log$_{10}$ ($F$$_{{\rm{}}}$)
    $= -17.64$}}. Completeness levels as a function of flux will
depend strongly on observed wavelength, and hold only for a particular
setup of detection software. We therefore only present these
limits as the ``average sample completeness'', to give a rough
indication of the depth of the LAE sample (particularly in Figure
\ref{fig:no_counts}, RHS, and {\mbox{ Figure \ref{fig:flux}}}).

\section{Results}
\label{sect:results}

\subsection{Luminosity Functions}
\label{res:lfs}

\begin{figure*}
\begin{center}
\resizebox{0.75\textwidth}{!}{\includegraphics{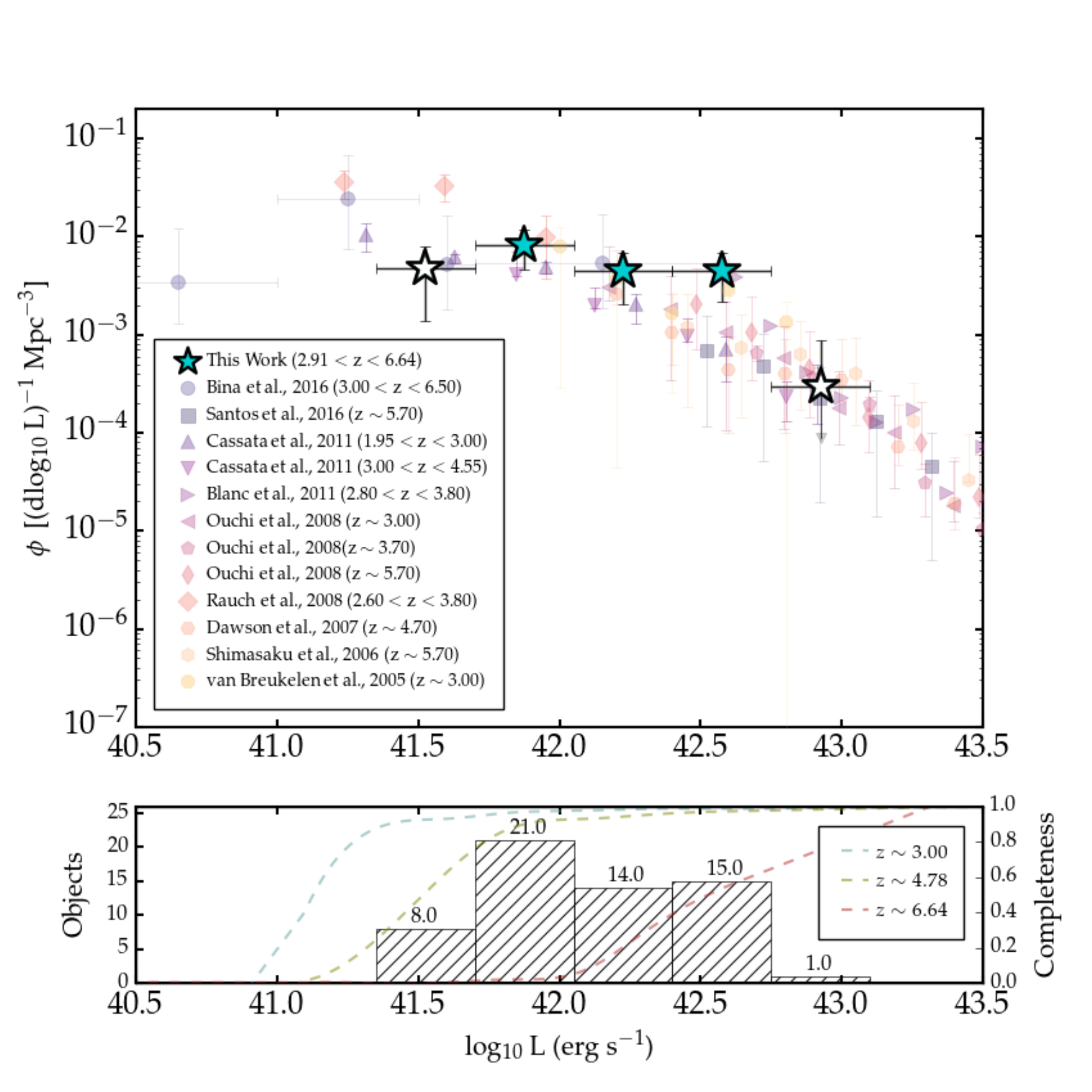}} 
\caption[]{In the upper panel we show the Ly$\alpha$ luminosity function estimated across the
  entire redshift range $2.91 < z < 6.64$ in the context of other
  surveys in the literature. We use Ly$\alpha$ fluxes estimated
  using a curve-of-growth analysis and calculate values of $\phi$
  using the $1/V_{max}$ estimator. Our faintest and brightest bins are marked with a
  transparent star to signify that the bins are incomplete
  and should be interpreted with caution. The literature data come
from narrow-band surveys (\citealt{Ouchi2008}: left-pointing
triangle at $z \approx 3.00$, pentagon at $z \approx 3.70$, narrow diamond
at $z \approx 5.70$, \citealt{Dawson2007}: hexagon at $z \approx 4.70$,
\citealt{Santos2016}: square at at $z \approx 5.70$),
blind or targetted long-slit spectroscopy (\citealt{Rauch2008}: wide diamond at $2.60 < z < 3.80$,
\citealt{Cassata2011}: upwards-pointing triangle at $1.95 < z < 3.00$,
downwards-pointing triangle at $3.00 < z < 4.55$,
\citealt{Shimasaku2006} at $ z \approx 5.70$), and IFU studies
(\citealt{vanBreukelen2005}; octagons at $z \approx 3.00$ and
\citealt{Blanc2011}; right-pointing triangle at $2.80 < z < 3.80$, \citealt{Bina2016};
circles at $3.00 < z < 6.50$). In the lower panel we show the
associated histogram of luminosities, overlaid with completeness
curves at three example redshifts. Our estimate of the luminosity function
sits higher than many literature
results in our most well-constrained bins, although within the $1\sigma$ Poissonian errorbars.}
\label{fig:LF}
\end{center}
\end{figure*}

Using the 59 objects presented here, we implement the
$1/V_{{\rm{max}}}$ estimator to assess the global luminosity
function for LAEs in the redshift range $2.91 < z < 6.64$. The results are presented in {\mbox{Table \ref{tab:objects}}} and
Figure \ref{fig:LF}. 

For each LAE, {\emph{i}}, in the catalogue, the
redshift $z_{i}$ is determined according to {{$z_{i} = {\lambda_{\;i}}/{1215.67} - 1.0$}}, where $\lambda_{\;i}$ is the observed wavelength of
Ly$\alpha$ according to the peak of the emission detected by {\sc
  muselet}. The luminosity $L_{i}$ is then computed according to
{{$L_i = f_{i} 4 \pi D_{L}^{2} (z_{i})$}}, where $f_{i}$ is the Ly$\alpha$ flux measured in our curve
of growth analysis, $D_{L}$ is the luminosity distance, and $z_{i}$ is the
Ly$\alpha$ redshift. The maximum co-moving volume within which this object could
be observed, $V_{{\rm{max}}}$($L_{\:i}$, $z_{\:i}$), is then computed by:

\begin{equation}
{V_{{\rm{max}}}(L_{i},\,z_{i}) = \int_{z1}^{z2} \frac{{\rm{d}}V}{{\rm{d}}z}\, {\rm{C}}(L_{i},z_{i})\, {\rm{d}}z}
\label{eq:Vmax}
\end{equation}

\noindent where $z_1 = 2.91$ and $z_2 = 6.64$, the minimum and maximum redshifts
of the survey respectively, d$V$ is the co-moving volume element corresponding to redshift interval
d$z = 0.01$, and C($L_{i}$, $z_{i}$) is the completeness curve for an
object of luminosity $L_i$, across all redshifts $z_i$.

The number density of objects per luminosity bin, {\rm{$\phi$}}, is then calculated according to:

\begin{equation}
{\phi [({\rm{d log_{10}}} L)^{-1} {\rm{Mpc}}^{-3}] = \sum_i \frac{1}{V_{\rm{max}}(L_{i}, z_{i})} / {\rm{bin size}}}
\label{eq:phi}
\end{equation}

\noindent where in this instance the bin size is $0.35$ dex. \\

\begin{table*}
\caption[]{Differential Ly$\alpha$ luminosity function in bins of
  $\Delta$ log$_{10}$ $L$= 0.35, including numbers of objects in each bin.}
\label{tab:objects}
\begin{center}
\begin{tabular}{cccr} \hline
 {\bf{Bin log$_{10}$ ($L$)}} [erg s$^{-1}$]& log$_{10}$ $L$$_{{\rm{median}}}$ [erg
 s$^{-1}$] & $\phi$ [(dlog$_{10}$ $L$)$^{-1}$ Mpc$^{-3}$]  & No.  \\
\hline
\hline
41.35 $<$ {\bf{41.525}} $<$ 41.70 & 41.596 & 0.0046 $\pm$ 0.0033 & 8\\
41.70 $<$ {\bf{41.875}} $<$ 42.05 & 41.872 & 0.0082 $\pm$ 0.0036 & 21\\
42.05 $<$ {\bf{42.225}} $<$ 42.40 & 42.247 & 0.0044 $\pm$ 0.0024 & 14\\
42.40 $<$ {\bf{42.575}} $<$ 42.75 & 42.508 & 0.0044 $\pm$ 0.0023 & 15\\
42.75 $<$ {\bf{42.925}} $<$ 43.10 & 42.829 & 0.0002 $\pm$ 0.0005 & 1\\


\hline
\end{tabular}
\end{center}
\end{table*} 

Figure \ref{fig:LF} shows the differential Ly$\alpha$ luminosity
function across the redshift range $2.91 < z < 6.64$, using a curve-of-growth analysis of the Ly$\alpha$ flux. In
the upper panel we show the values of $\phi$ given by the $1/V_{max}$
estimator, and in the lower panel a histogram depicts the
numbers of objects found in each bin. Overlaid on the lower panels are
the completeness curves as a function of luminosity at three example
redshifts ($z = 3.00, 4.78$ and $6.64$) to give
an indication of the range of completeness corrections being applied to
objects in each bin. Notably in the lowest luminosity bins
completeness corrections can range from $0$ to over $90$ percent, and so
small inaccuracies in the completeness estimate will potentially
result in significant changes to the luminosity function here.

In our central luminosity range where the bins are most well-populated,
our data are a factor of $2-3$ higher than many of the results from
previous studies, although the $1\sigma$ Poissonian error on our
points touches a couple of the literature results, or failing this the
error bars overlap. Additionally, although our highest luminosity bin
(containing only a single object) is in
perfect agreement with the well-constrained literature at this luminosity, we note that
the bin is incomplete, and correcting for this would likely place the
data point above the literature again.

\subsection{Comparison to Literature}
\label{sect:literature}

In the following paragraphs we make a more detailed comparison to
literature results from the various commonly-adopted approaches to LAE
selection: narrow-band studies, blind long-slit spectroscopy, early
IFU data, and lensed LAEs detected with MUSE \citep{Bina2016}\footnote{Note that the points
we show here from Abell $1689$ are an updated version of
\cite{Bina2016} correcting for an error in the survey volume that they
originally calculated.}. We note
that these studies themselves are dispersed due to cosmic variance,
slit-losses, small apertures and different equivalent width limits. 

 In comparison to the narrow-band studies shown, our data sit higher
 at all luminosities than \cite{Ouchi2008} at $z =
 3.00$ and $z = 3.70$, but always within the
 error bars of the narrowband data. We are in agreement across all
 datapoints with \cite{Ouchi2008} at $z =
 5.00$, however this study does have larger error bars than their lower redshift data. The final two narrowband studies are
 \cite{Dawson2007} and \cite{Santos2016} at $z = 4.70$ and $z = 5.70$
 respectively, and fully consistent
 with one another. The \cite{Dawson2007}  data reach
 {\mbox{log$_{10}$($L$$_{{\rm{}}}$) $ = 42.0$}} and are in good
 agreement with our points except for our bin at
 {\mbox{log$_{10}$($L$$_{{\rm{}}}$) $ = 42.6$}} where our value
 of $\phi$ is significantly higher. The \cite{Santos2016} data are
 similar, they reach down to slightly brighter than  {\mbox{log$_{10}$($L$$_{{\rm{}}}$) $ = 42.5$}}, and the only
 point in disagreement with our own is again our measurement at  {\mbox{log$_{10}$($L$$_{{\rm{}}}$) $ = 42.6$}} where
 we are significantly higher. 

Our data point at  {\mbox{log$_{10}$($L$$_{{\rm{}}}$) $ = 42.6$}} sits almost exactly on top of two other data
points, coming from the two other IFU studies we examine
(\citealt{vanBreukelen2005}, \citealt{Blanc2011}). For \cite{Blanc2011}
this is the faintest datapoint in their sample, but
\cite{vanBreukelen2005} reach almost 1 dex deeper where the data
points agree with our data and are more in line with the rest of the
literature as well. The error bars of the three IFU data points overlap
with those of all three datasets from \cite{Ouchi2008}, but it is interesting to note that
all three studies are high in the
{\mbox{log$_{10}$($L$$_{{\rm{}}}$) $ = 42.6$}} bin, and
inconsistent with both \cite{Dawson2007} and \cite{Santos2016}.

Our faintest two bins at  {\mbox{log$_{10}$($L$$_{{\rm{}}}$) $ = 41.75$}} and  {\mbox{log$_{10}$($L$$_{{\rm{}}}$) $ = 41.5$}} can only be compared to the
deep blind long-slit spectroscopy of \cite{Cassata2011} and
\cite{Rauch2008}, and the MUSE results from the lensing cluster Abell
$1689$ \citep{Bina2016}. Our  {\mbox{log$_{10}$($L$$_{{\rm{}}}$) $ = 41.75$}} datapoint sits between the
brightest point from \cite{Rauch2008} and the faintest point from
\cite{Cassata2011}  at redshift $3.00 < z < 4.55$. Our value of $\phi$
is consistent with both these points. Our lowest luminosity point lies
below \cite{Cassata2011} at $1.95 < z < 3.00$ although consistent with
their result within errors, however the point lies significantly below
the measurement of \cite{Rauch2008}. Since this point is our lowest
luminosity point, and is likely to suffer the most from small
inaccuracies in completeness estimates, we do not interpret this point
as having ruled out the result of \cite{Rauch2008}, although the data
are noticeably in better agreement with the values of
\cite{Cassata2011}. 
Finally, we compare to the results of \cite{Bina2016} who calculated the number
density of LAEs behind lensing cluster
Abell $1689$. The $17$ LAE luminosities range between $40.5 < {\rm{log_{10} L_{}}}
< 42.5$ erg s$^{-1}$, i.e. the deepest LAE data to date, and despite
having applied no completeness correction, the values are broadly consistent with our own estimates. The errors on the small
number of objects in \cite{Bina2016}  mean that their values of $\phi$
are entirely consistent with \cite{Cassata2011}, as well as close to
consistent with the \cite{Rauch2008} data points. 

\subsection{Cumulative Luminosity Function}
\label{sect:cumul}
\begin{figure}
\begin{center}
 \includegraphics[width=0.48\textwidth]{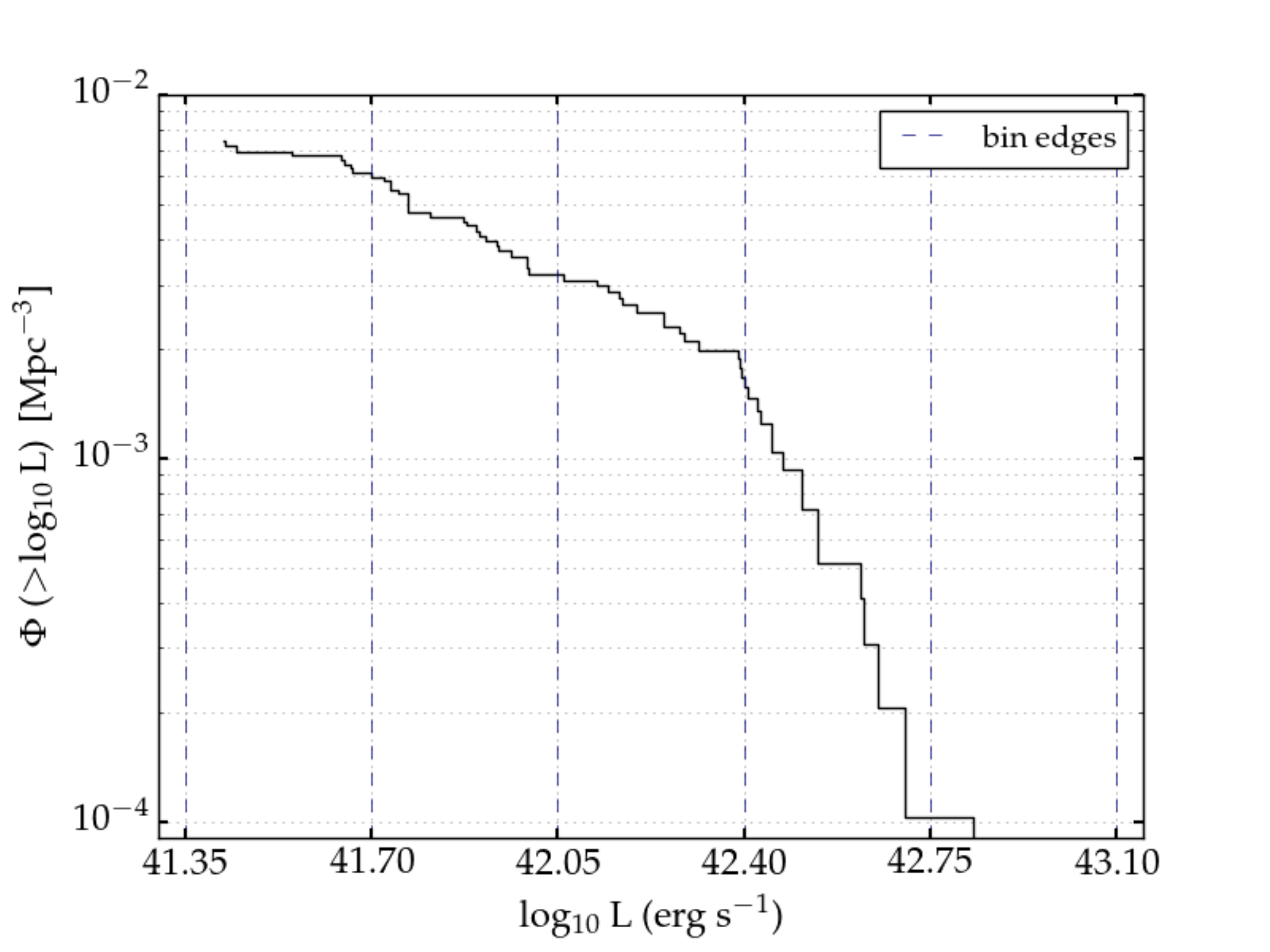} 
\caption[]{The cumulative Ly$\alpha$ luminosity function estimated across the
  entire redshift range $2.91 < z < 6.64$. Note that for
  clarity we zoom-in on the range of the data (this figure only). The step function gives an
  indication of how objects' luminosities are distributed within each
  bin of the luminosity function in Figure \ref{fig:LF}, and allows us
  to interpret each bin with the appropriate level of caution.}
\label{fig:LF_cum}
\end{center}
\end{figure}

In Figure \ref{fig:LF_cum} we show the cumulative Ly$\alpha$
luminosity function between redshifts $2.91 < z < 6.64$. Note that we
show this Figure in a ``zoomed-in'' panel due to the data covering
only a small part of the dynamic range of the literature shown in
Figure \ref{fig:LF}. The cumulative luminosity
function has the advantage of being sensitive to each individual
object in the sample, alleviating the problem of lost information in
a binned differential luminosity function, especially for small samples. Each object in the sample
is visible in the function as a vertical step at the luminosity of the
object. This allows us to visualise the distribution of objects
within each bin of the differential luminosity function in Figure
\ref{fig:LF}, for instance our lowest luminosity bin contains two
objects towards the lower limit but is by no means an even
distribution of luminosities. Equally, our highest luminosity bin
contains only a single object lying towards lower luminosity edge. This allows us to interpret the density
estimates in these bins with the appropriate level of caution.

\subsection{Comparison to Models}
\label{sect:models}
\begin{figure}
\begin{center}
 \includegraphics[width=0.48\textwidth]{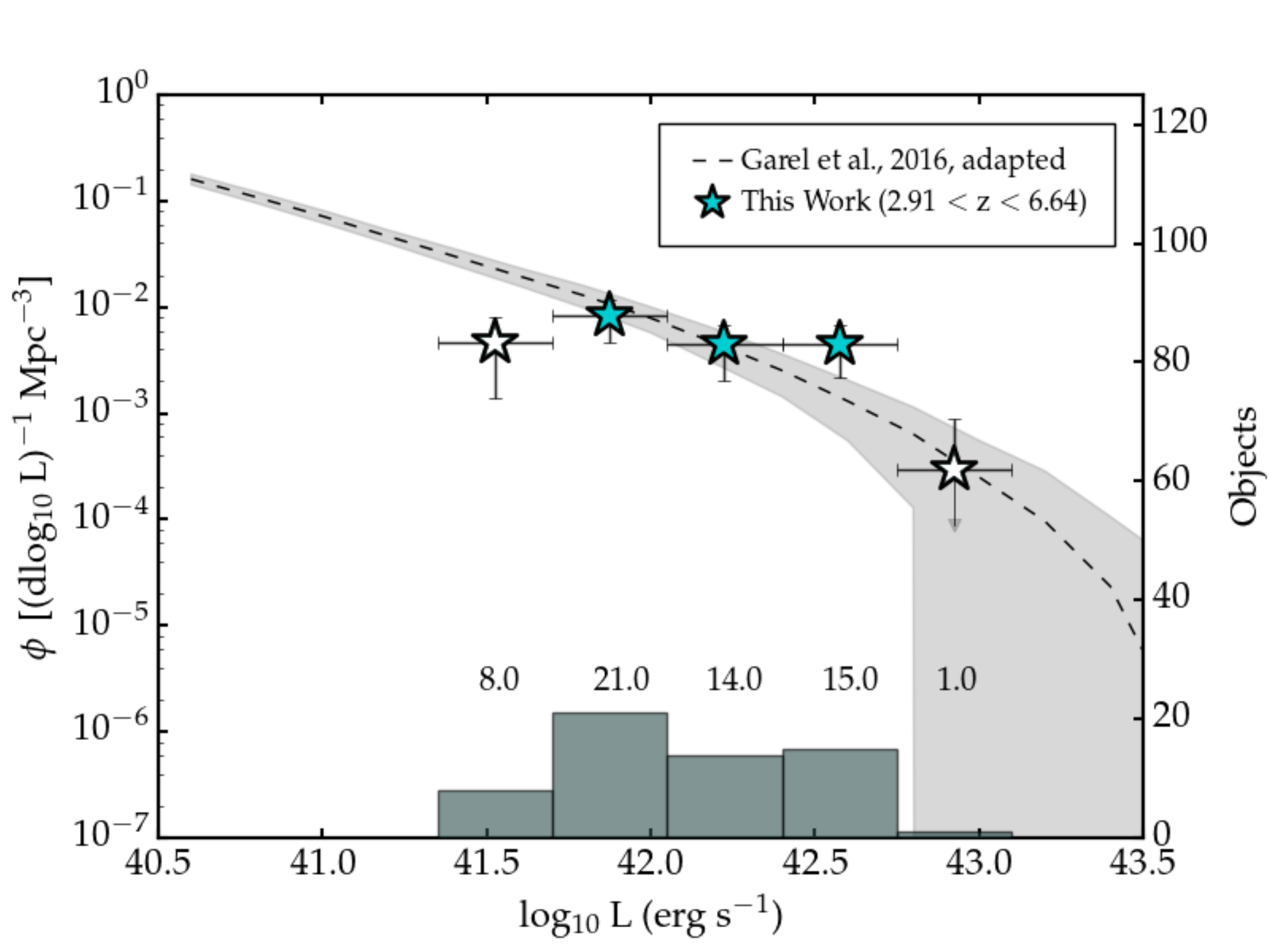} 
 \caption{Comparison of our differential luminosity function to
   predictions from the semi-analytic model of Garel et al. 2015
   (similar to those shown in Garel et al. 2016 but adapted to our
   survey volume). The error on the model prediction
   is the standard deviation from 1000 realisations of the mock
   lightcones produced by the model of Garel et al. (2015). The major
   component of this scatter comes from relative cosmic variance
   defined as the scatter in excess of that predicted by Poisson shot noise.}
  \label{fig:model}
\end{center}
\end{figure}

In Figure \ref{fig:model} we compare our results to predictions from
the mock lightcones used in \cite{Garel2016}. The lightcones are generated
using the model presented in \citealt{Garel2015} (see \citealt{Garel2012}
for details) whereby the
GALICS hybrid model of galaxy formation \citep{Hatton2003} is
coupled with numerical simulations of Ly$\alpha$ radiative transfer. GALICS combines an N-body cosmological
simulation to follow the hierarchical growth of dark matter
structures in a representative co-moving volume of (100 h$^{-1}$ Mpc)$^3$ with a semi-analytic component to describe the evolution of
baryons within virialised dark matter halos. The escape of Ly$\alpha$
photons occurs through galactic outflows modelled as thin expanding
shells of gas and dust (\citealt{Schaerer2011},
\citealt{Verhamme2006}). The escape
fraction of each galaxy  in \cite{Garel2015} is then determined by
interpolating the shell parameters (expansion speed, HI column density, dust opacity and velocity dispersion) predicted by GALICS onto the grid of
radiative transfer models of \cite{Schaerer2011}.

We show the mean luminosity function over 1000 realisations of mock lightcones computed
with the model of \cite{Garel2015}. The geometry of the lightcones is adapted
to mimic our survey of the HDFS  (i.e. $2.91 < z < 6.64$ over 1
square arcminute) and error bars give the 1$\sigma$ standard deviation of the
measurement. Our data are in good agreement with the model, lying
directly on top of the predicted number densities in the  {\mbox{log$_{10}$($L$$_{{\rm{}}}$) $ = 41.88$}} and
 {\mbox{log$_{10}$($L$$_{{\rm{}}}$) $ = 42.23$}} bins as well as the single object in our  {\mbox{log$_{10}$($L$$_{{\rm{}}}$) $ = 42.93$}} bin.
Our measurement in the {\mbox{log$_{10}$($L$$_{{\rm{}}}$) $ = 42.58$}}
bin once again sits high by a factor of $\approx 3$,
with a $1\sigma$ error bar just touching the $1\sigma$ error on the model predictions. Finally, our lowest luminosity point at  {\mbox{log$_{10}$($L$$_{{\rm{}}}$) $ = 41.53$}} falls
well below the model predictions from \cite{Garel2015}, which is not
surprising given the incomplete sampling of the bin. In the model,
low-SFR galaxies have a higher Ly$\alpha$ escape fraction due to lower
gas and dust contents such that the luminosity function continues to rise
steeply towards faint luminosities ($< 10^{42}$ erg s$^{-1}$). This again
emphasises the need for a more sophisticated completeness assessment
for our sample to derive a more robust estimate of the luminosity function at the faint end, and thus better constrain LAE models.

Additionally, \cite{Garel2016} showed that the uncertainty on number counts in
surveys of this volume will be dominated
by cosmic variance, defined as the uncertainty in excess to
that predicted by Poisson shot noise. This adds further motivation to our 9 by 9 square
arcminute observations of the HUDF field with MUSE. In this
forthcoming study, cosmic variance will be
substantially reduced, but our data will still be deep enough to probe
well below the knee of the luminosity function.

\section{Discussion}
\label{sect:disc}

\subsection{Luminosity Functions from Different Flux Estimates}
\label{sect:LF_flux}

\begin{figure}
\begin{center}
 \includegraphics[width=0.48\textwidth]{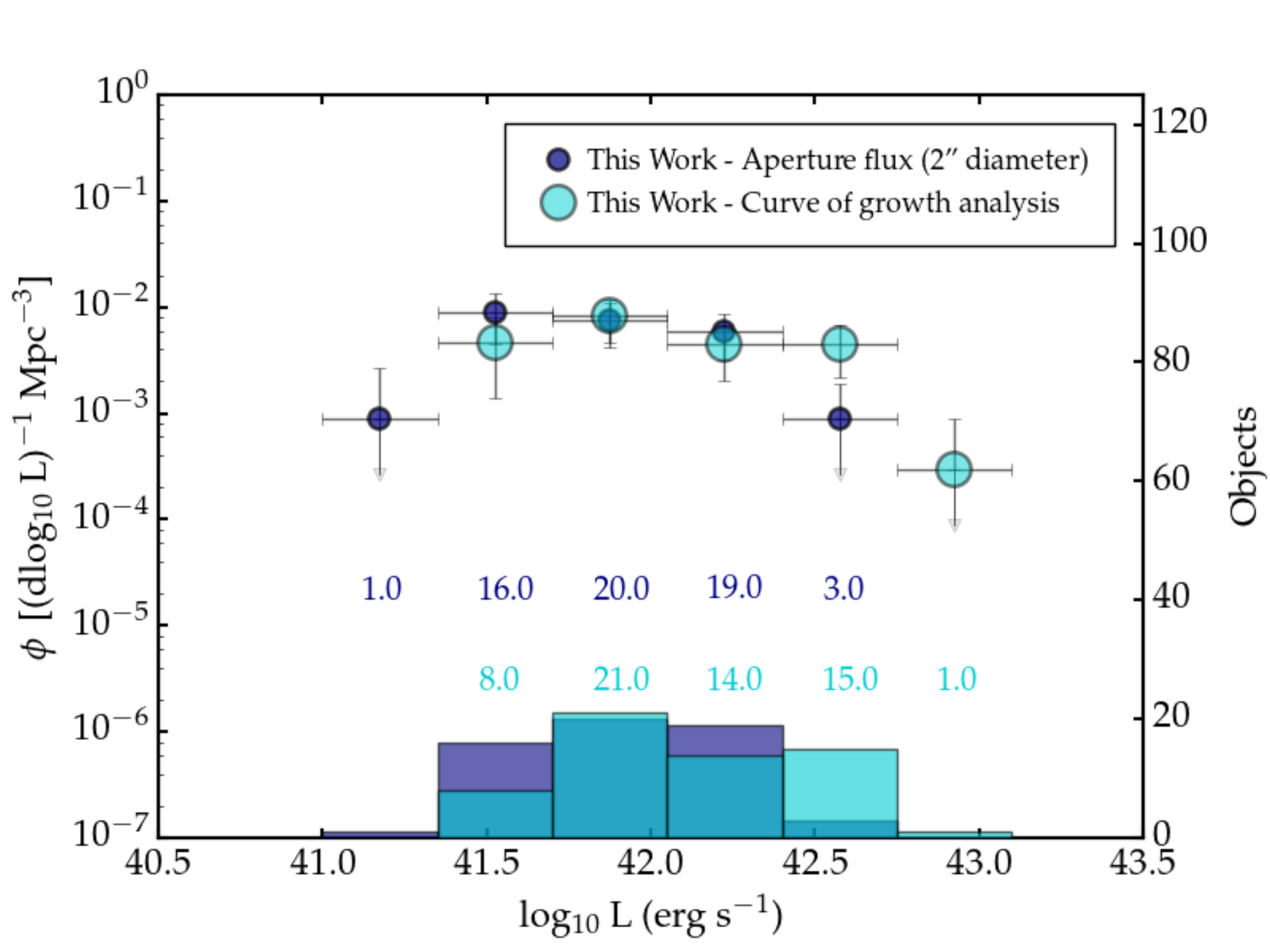} 
 \caption{Luminosity functions from two different methods of estimating the total LAE flux. The
   small dark blue circles give values of $\phi$ for photometry from a 2-arcsecond
   diameter aperture, and large light blue circles show values of
   $\phi$ from a curve of growth analysis of total LAE flux. Due to
   objects shifting between bins the central measurements are in
   agreement, but the two
   approaches give different impressions of the luminosity range being
 studied, and will make a significant difference to measurements of
 the faint-end slope.}
  \label{fig:LFs_flux}
\end{center}
\end{figure}

In Section \ref{sect:flux} we highlighted the difficulties of flux
estimation from long-slit spectroscopy, in addition to the problems that
arise from small aperture photometry. Figure \ref{fig:flux}
demonstrates how an aperture of diameter 2 arcseconds misses a great
deal of flux for LAEs, which are known to exhibit extended
emission. Here, in Figure \ref{fig:LFs_flux}, we
examine the effect that each of these approaches to the flux measurement has on our resultant
luminosity function. The luminosity function from $F_{{\rm{\;C.o.G}}}$ (shown in
the large light blue circles) is the same that we show in Figure
\ref{fig:LF}, and the luminosity function from $F_{{\rm{\;2\arcsec}}}$ is shown on the same axes in smaller dark blue
circles. Changing the method of flux
estimation means that objects jump between bins giving a different
impression of the luminosity range under study. In the brightest overlap bin at
{\mbox{log$_{10}$($L_{{\rm{}}}$) $ = 42.57$}}, the value of
$\phi$($F_{{\rm{\;C.o.G}}}$) is significantly above $\phi$($F_{{\rm{\;2\arcsec}}}$) as the measured flux of many objects has increased. In the
faintest overlap bin however the opposite effect is seen, since some
objects have shifted out of the bin towards higher measured
luminosities. Notably the value of $\phi$($F_{{\rm{\;2\arcsec}}}$ in this bin
is in very good agreement with the value found in most literature
studies. Realistic flux estimates will be of even greater
importance when it comes to assessing the faint end slope, and so we
re-assert their importance here.

\subsection{Test for Evolution}
\label{res:evolution}
\begin{figure}
\begin{center}
\resizebox{0.48\textwidth}{!}{\includegraphics{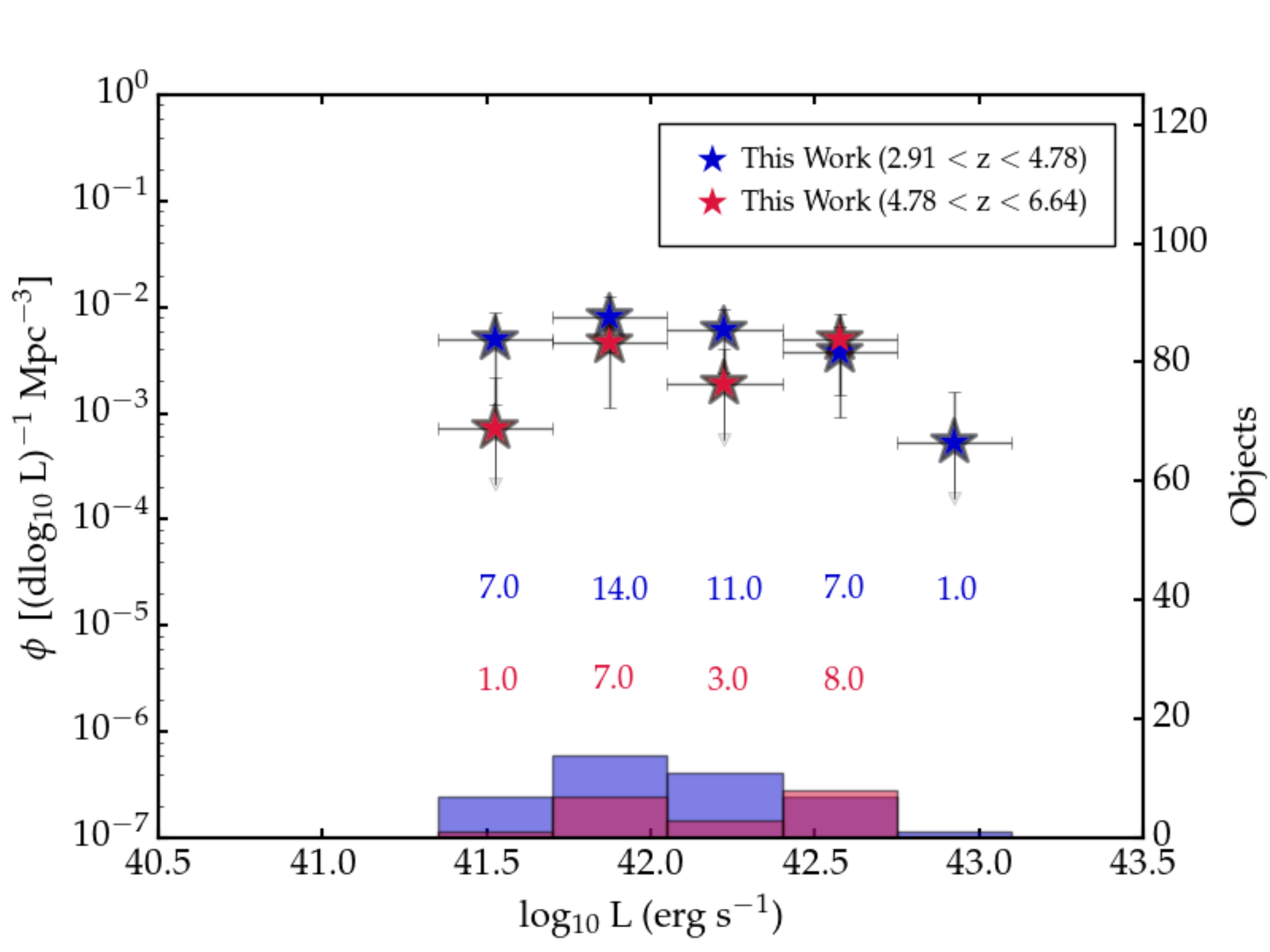}} 
\caption[]{Luminosity functions derived for
the high and low redshift halves of the dataset when split at the central LAE redshift detectable with MUSE. Blue stars show the low
redshift half of the data, and red stars show the high redshift half
of the data. At the bottom of the panel we show histograms of the luminosity
distributions of the two halves of the dataset. The two distributions
lie within the errorbars of one another, giving no evidence to suggest
any redshift evolution in the observed luminosity function.}
\label{fig:LF_comparison}
\end{center}
\end{figure}

Many studies have looked for signs of redshift-evolution in the observed
Ly$\alpha$ luminosity function. \cite{vanBreukelen2005},
\cite{Shimasaku2006}, \cite{Ouchi2008} and \cite{Cassata2011} all 
concluded that there was no evidence of such evolution in their
data. Here, with our small sample of objects, we can only make a crude
attempt to look for evolution between $z = 6.64$ and $z = 2.91$. We split the sample into high- and low-redshift subsets at the
centre of the LAE redshift range and compare the two halves of the data, the number densities of objects
in the two subsets can be seen in
Figure \ref{fig:LF_comparison}. We see very little difference
between the two halves, and indeed in each of the luminosity bins
populated by both samples, the values of $\phi$ are within the error
bars of one another. As an additional check, we use a two-sample Kolmogorov-Smirnov test on the
two distributions of volume-corrected
luminosities (i.e. the values of 1/V$_{{\rm{max}}}$) and find a
two-tailed p-value of $p= 3.345 \times 10^{-9}$ meaning we
cannot discount the null hypothesis that the two distributions were drawn from
the same underlying population. We conclude that there is no evidence
of strong evolution between the two halves of the data --
consistent with literature results -- but note that we are limited by the small numbers
of objects here, and a re-examination of the question is warranted with a richer dataset (Drake et al., in prep).

\subsection{Limitations of our Study}
\label{res:limitations}
Clearly our interpretation of the luminosity function is restricted by the small
number of objects presented here, and the limitations of the
$1/V_{max}$ estimator. Although the faint-end of our luminosity
function is broadly consistent with previous studies, our sample is
not rich enough to constrain the steepness of the slope. In Drake et al., (in prep) we will
dramatically increase the size of our dataset using LAEs in the {\emph{Hubble
Ultra Deep Field}} (HUDF) including
several hundred sources from the MUSE HUDF $9 \times 9$ square
arcminute ``mosaic'' field. As discussed in Section \ref{sect:models},
\cite{Garel2016} demonstrated that a study of this size and depth is
the ideal survey design to examine the bulk of LAEs, while minimising
the contribution of cosmic variance. In two complementary studies, the MUSE-WIDE program will
substantially beat down statistics at the bright-end of the luminosity
function (Herenz et al., in prep), and MUSE GTO lensing fields will provide more of the deepest samples of LAEs to date. The combination of the
MUSE-WIDE, MUSE-DEEP and MUSE-lensing fields will allow constraints on
both the bright- and faint-end of the luminosity function,
resulting in the most accurate
measurement of the faint-end slope to date.

\section{Conclusions}
\label{sect:concl}
In summary, we have presented a homogeneous, automatically-detected sample of $59$ LAEs in the HDFS using blind
spectroscopy from MUSE. We validate the Ly$\alpha$ line through a
careful matching to the deeper (heterogeneously constructed) catalogue of
B15. We have shown that
the method of Ly$\alpha$ flux estimation can significantly alter
measured Ly$\alpha$ fluxes and investigated the effect this has on the
luminosity function. We have designed
a procedure self-consistent with our detection software to determine
our selection function through recovery of fake point-source line
emitters from deep MUSE datacubes, and compute a
global Ly$\alpha$ luminosity function using a curve-of-growth
analysis of the Ly$\alpha$ flux, and implementing the $1/V_{max}$
estimator. We compare our results to literature studies, and semi-analytic model predictions from
\cite{Garel2015}, before finally examining the dataset for signs of evolution in the observed
luminosity function.\\

Our main conclusions can be broadly summarised as follows:

\begin{itemize}
\renewcommand{\labelitemi}{$\bullet$}

\item We automatically detect 59 LAEs in the HDFS across a
  flux range of  $\approx -18.0 < {\rm{log\; F}} < -16.3\; ({\rm{erg
      s}}^{-1} {\rm{cm}}^{-2})$ using homogeneous and robust selection
  criteria, validating each LAE by matching to the deep catalogue of B15.

\item Our global luminosity function between $2.91 < z < 6.64$ sits
  higher by a factor of $2-3$ than the literature in our most well-constrained bins, although
  $1\sigma$ errorbars overlap with the data of several literature
  studies at the same luminosity. 

\item The small drop in number density
  between our penultimate and faintest luminosity bin is likely to be entirely
  due to the limitations of our method; namely the effect of
  incomplete bins on the $1/V_{max}$ estimator, and our idealised completeness assessment where LAEs are treated
  as point sources. We will investigate this in Drake et al., (in
  prep) using the MUSE HUDF mosaic sample.

\item Our luminosity function is in good agreement with the
  semi-analytical model of \cite{Garel2015} with the exception of our
  bin at log$_{10} \,L = 42.58$. The bin is once again a factor $\approx 3$ higher than
  the predictions, with a $1\sigma$ Poissonian error bar that just
  touches the $1\sigma$ error on the model predictions. 

\item Method of Ly$\alpha$ flux estimation plays a role in the
determination of the Ly$\alpha$ luminosity function and becomes most significant when
  measuring the faint end slope. Care should be
  taken here as studies start to probe further into the low-luminosity
  LAE population.

\item When splitting our data at the central redshift and comparing
  the two halves of the data we see no evidence for strong evolution in the observed Ly$\alpha$
  luminosity function across the redshift range $2.91 < z <
  6.64$. This is entirely consistent with results
  in the literature. 
\end{itemize}

Our pilot study demonstrates the efficiency of MUSE as a detection
machine for emission-line galaxies, and strongly motivates our
analysis of the HUDF  $9 \times 9$ square
arcminute mosaic. The conservative nature of our selection process means that the objects presented here
represent a robustly-selected sub-sample of the galaxies MUSE will
ultimately detect and identify, and this is very encouraging for the
potential of additional blank-field datasets from MUSE.

\section*{Acknowledgements}
This work has been carried out thanks to the support of the ANR FOGHAR
(ANR-13-BS05-0010-02), the OCEVU Labex (ANR-11-LABX-0060) and the
A*MIDEX project (ANR- 11-IDEX-0001-02) funded by the ``Investissements
d'avenir'' French government program managed by the ANR. 
ABD
thanks CRAL and ANR FOGHAR (ANR-13-BS05-0010-02), the Astrophysics Research
Institute, Liverpool John Moores University, UK, for supporting a
visiting fellowship, and Phil James for proof-reading. We make extensive use of the Muse Python Data
Analysis Framework software ({\sc mpdaf}), thanks to L.Piqueras. JR
acknowledges support from the ERC starting grant 336736-CALENDS. TG is
grateful to the LABEX Lyon Institute of Origins (ANR-10-LABX-0066) of
the Universit\'{e} de Lyon for its financial support within the
programme 'Investissements d'Avenir' (ANR-11-IDEX-0007) of the French
government operated by the National Research Agency (ANR). LW and ECH
acknowledge support by the Competitive Fund of the Leibniz Association
through grant SAW-2015-AIP-2. JS acknowledges
278594-GasAroundGalaxies. 

\bibliographystyle{mnras}
\bibliography{paper}

\begin{thebibliography}{}
\makeatletter
\relax
\def\mn@urlcharsother{\let\do\@makeother \do\$\do\&\do\#\do\^\do\_\do\%\do\~}
\def\mn@doi{\begingroup\mn@urlcharsother \@ifnextchar [ {\mn@doi@}
  {\mn@doi@[]}}
\def\mn@doi@[#1]#2{\def\@tempa{#1}\ifx\@tempa\@empty \href
  {http://dx.doi.org/#2} {doi:#2}\else \href {http://dx.doi.org/#2} {#1}\fi
  \endgroup}
\def\mn@eprint#1#2{\mn@eprint@#1:#2::\@nil}
\def\mn@eprint@arXiv#1{\href {http://arxiv.org/abs/#1} {{\tt arXiv:#1}}}
\def\mn@eprint@dblp#1{\href {http://dblp.uni-trier.de/rec/bibtex/#1.xml}
  {dblp:#1}}
\def\mn@eprint@#1:#2:#3:#4\@nil{\def\@tempa {#1}\def\@tempb {#2}\def\@tempc
  {#3}\ifx \@tempc \@empty \let \@tempc \@tempb \let \@tempb \@tempa \fi \ifx
  \@tempb \@empty \def\@tempb {arXiv}\fi \@ifundefined
  {mn@eprint@\@tempb}{\@tempb:\@tempc}{\expandafter \expandafter \csname
  mn@eprint@\@tempb\endcsname \expandafter{\@tempc}}}

\bibitem[\protect\citeauthoryear{{Bacon} et~al.,}{{Bacon}
  et~al.}{2010}]{Bacon2010}
{Bacon} R.,  et~al., 2010, in Ground-based and Airborne Instrumentation for
  Astronomy III. p. 773508, \mn@doi{10.1117/12.856027}

\bibitem[\protect\citeauthoryear{{Bacon} et~al.,}{{Bacon}
  et~al.}{2015}]{Bacon2015}
{Bacon} R.,  et~al., 2015, \mn@doi [\aap] {10.1051/0004-6361/201425419}, \href
  {http://adsabs.harvard.edu/abs/2015A%26A...575A..75B} {575, A75}

\bibitem[\protect\citeauthoryear{{Bertin} \& {Arnouts}}{{Bertin} \&
  {Arnouts}}{1996}]{BertinArnouts1996}
{Bertin} E.,  {Arnouts} S.,  1996, \mn@doi [\aaps] {10.1051/aas:1996164}, \href
  {http://adsabs.harvard.edu/abs/1996A%26AS..117..393B} {117, 393}

\bibitem[\protect\citeauthoryear{{Bina} et~al.,}{{Bina}
  et~al.}{2016}]{Bina2016}
{Bina} D.,  et~al., 2016, \mn@doi [\aap] {10.1051/0004-6361/201527913}, \href
  {http://adsabs.harvard.edu/abs/2016A%26A...590A..14B} {590, A14}

\bibitem[\protect\citeauthoryear{{Blanc} et~al.,}{{Blanc}
  et~al.}{2011}]{Blanc2011}
{Blanc} G.~A.,  et~al., 2011, \mn@doi [\apj] {10.1088/0004-637X/736/1/31},
  \href {http://adsabs.harvard.edu/abs/2011ApJ...736...31B} {736, 31}

\bibitem[\protect\citeauthoryear{{Borisova} et~al.,}{{Borisova}
  et~al.}{2016}]{Borisova2016}
{Borisova} E.,  et~al., 2016, preprint, \href
  {http://adsabs.harvard.edu/abs/2016arXiv160501422B} {} (\mn@eprint {arXiv}
  {1605.01422})

\bibitem[\protect\citeauthoryear{{Casertano} et~al.,}{{Casertano}
  et~al.}{2000}]{Casertano2000}
{Casertano} S.,  et~al., 2000, \mn@doi [\aj] {10.1086/316851}, \href
  {http://adsabs.harvard.edu/abs/2000AJ....120.2747C} {120, 2747}

\bibitem[\protect\citeauthoryear{{Cassata} et~al.,}{{Cassata}
  et~al.}{2011}]{Cassata2011}
{Cassata} P.,  et~al., 2011, \mn@doi [\aap] {10.1051/0004-6361/201014410},
  \href {http://adsabs.harvard.edu/abs/2011A%26A...525A.143C} {525, A143}

\bibitem[\protect\citeauthoryear{{Cowie} \& {Hu}}{{Cowie} \&
  {Hu}}{1998}]{Cowie&Hu1998}
{Cowie} L.~L.,  {Hu} E.~M.,  1998, \mn@doi [\aj] {10.1086/300309}, \href
  {http://adsabs.harvard.edu/abs/1998AJ....115.1319C} {115, 1319}

\bibitem[\protect\citeauthoryear{{Dawson}, {Rhoads}, {Malhotra}, {Stern},
  {Wang}, {Dey}, {Spinrad}  \& {Jannuzi}}{{Dawson} et~al.}{2007}]{Dawson2007}
{Dawson} S.,  {Rhoads} J.~E.,  {Malhotra} S.,  {Stern} D.,  {Wang} J.,  {Dey}
  A.,  {Spinrad} H.,   {Jannuzi} B.~T.,  2007, \mn@doi [\apj] {10.1086/522908},
  \href {http://adsabs.harvard.edu/abs/2007ApJ...671.1227D} {671, 1227}

\bibitem[\protect\citeauthoryear{{Dayal} \& {Libeskind}}{{Dayal} \&
  {Libeskind}}{2012}]{Dayal2012}
{Dayal} P.,  {Libeskind} N.~I.,  2012, \mn@doi [\mnras]
  {10.1111/j.1745-3933.2011.01166.x}, \href
  {http://adsabs.harvard.edu/abs/2012MNRAS.419L...9D} {419, L9}

\bibitem[\protect\citeauthoryear{{Deharveng} et~al.,}{{Deharveng}
  et~al.}{2008}]{Deharveng2008}
{Deharveng} J.-M.,  et~al., 2008, \mn@doi [\apj] {10.1086/587953}, \href
  {http://adsabs.harvard.edu/abs/2008ApJ...680.1072D} {680, 1072}

\bibitem[\protect\citeauthoryear{{Dijkstra}, {Gronke}  \&
  {Venkatesan}}{{Dijkstra} et~al.}{2016}]{Dijkstra2016}
{Dijkstra} M.,  {Gronke} M.,   {Venkatesan} A.,  2016, preprint, \href
  {http://adsabs.harvard.edu/abs/2016arXiv160408208D} {} (\mn@eprint {arXiv}
  {1604.08208})

\bibitem[\protect\citeauthoryear{{Drake} et~al.,}{{Drake}
  et~al.}{2013}]{Drake2013}
{Drake} A.~B.,  et~al., 2013, \mn@doi [\mnras] {10.1093/mnras/stt775}, \href
  {http://adsabs.harvard.edu/abs/2013MNRAS.433..796D} {433, 796}

\bibitem[\protect\citeauthoryear{{Drake} et~al.,}{{Drake}
  et~al.}{2015}]{Drake2015}
{Drake} A.~B.,  et~al., 2015, \mn@doi [\mnras] {10.1093/mnras/stv2027}, \href
  {http://adsabs.harvard.edu/abs/2015MNRAS.454.2015D} {454, 2015}

\bibitem[\protect\citeauthoryear{{Dressler}, {Martin}, {Henry}, {Sawicki}  \&
  {McCarthy}}{{Dressler} et~al.}{2011}]{Dressler2011}
{Dressler} A.,  {Martin} C.~L.,  {Henry} A.,  {Sawicki} M.,   {McCarthy} P.,
  2011, \mn@doi [\apj] {10.1088/0004-637X/740/2/71}, \href
  {http://adsabs.harvard.edu/abs/2011ApJ...740...71D} {740, 71}

\bibitem[\protect\citeauthoryear{{Dressler}, {Henry}, {Martin}, {Sawicki},
  {McCarthy}  \& {Villaneuva}}{{Dressler} et~al.}{2015}]{Dressler2015}
{Dressler} A.,  {Henry} A.,  {Martin} C.~L.,  {Sawicki} M.,  {McCarthy} P.,
  {Villaneuva} E.,  2015, \mn@doi [\apj] {10.1088/0004-637X/806/1/19}, \href
  {http://adsabs.harvard.edu/abs/2015ApJ...806...19D} {806, 19}

\bibitem[\protect\citeauthoryear{{Garel}, {Blaizot}, {Guiderdoni}, {Schaerer},
  {Verhamme}  \& {Hayes}}{{Garel} et~al.}{2012}]{Garel2012}
{Garel} T.,  {Blaizot} J.,  {Guiderdoni} B.,  {Schaerer} D.,  {Verhamme} A.,
  {Hayes} M.,  2012, \mn@doi [\mnras] {10.1111/j.1365-2966.2012.20607.x}, \href
  {http://adsabs.harvard.edu/abs/2012MNRAS.422..310G} {422, 310}

\bibitem[\protect\citeauthoryear{{Garel}, {Blaizot}, {Guiderdoni},
  {Michel-Dansac}, {Hayes}  \& {Verhamme}}{{Garel} et~al.}{2015}]{Garel2015}
{Garel} T.,  {Blaizot} J.,  {Guiderdoni} B.,  {Michel-Dansac} L.,  {Hayes} M.,
   {Verhamme} A.,  2015, \mn@doi [\mnras] {10.1093/mnras/stv374}, \href
  {http://adsabs.harvard.edu/abs/2015MNRAS.450.1279G} {450, 1279}

\bibitem[\protect\citeauthoryear{{Garel}, {Guiderdoni}  \& {Blaizot}}{{Garel}
  et~al.}{2016}]{Garel2016}
{Garel} T.,  {Guiderdoni} B.,   {Blaizot} J.,  2016, \mn@doi [\mnras]
  {10.1093/mnras/stv2467}, \href
  {http://adsabs.harvard.edu/abs/2016MNRAS.455.3436G} {455, 3436}

\bibitem[\protect\citeauthoryear{{Gronke}, {Dijkstra}, {Trenti}  \&
  {Wyithe}}{{Gronke} et~al.}{2015a}]{Gronke2015a}
{Gronke} M.,  {Dijkstra} M.,  {Trenti} M.,   {Wyithe} S.,  2015a, \mn@doi
  [\mnras] {10.1093/mnras/stv329}, \href
  {http://adsabs.harvard.edu/abs/2015MNRAS.449.1284G} {449, 1284}

\bibitem[\protect\citeauthoryear{{Gronke}, {Bull}  \& {Dijkstra}}{{Gronke}
  et~al.}{2015b}]{Gronke2015b}
{Gronke} M.,  {Bull} P.,   {Dijkstra} M.,  2015b, \mn@doi [\apj]
  {10.1088/0004-637X/812/2/123}, \href
  {http://adsabs.harvard.edu/abs/2015ApJ...812..123G} {812, 123}

\bibitem[\protect\citeauthoryear{{Hatton}, {Devriendt}, {Ninin}, {Bouchet},
  {Guiderdoni}  \& {Vibert}}{{Hatton} et~al.}{2003}]{Hatton2003}
{Hatton} S.,  {Devriendt} J.~E.~G.,  {Ninin} S.,  {Bouchet} F.~R.,
  {Guiderdoni} B.,   {Vibert} D.,  2003, \mn@doi [\mnras]
  {10.1046/j.1365-8711.2003.05589.x}, \href
  {http://adsabs.harvard.edu/abs/2003MNRAS.343...75H} {343, 75}

\bibitem[\protect\citeauthoryear{{Henry}, {Martin}, {Dressler}, {Sawicki}  \&
  {McCarthy}}{{Henry} et~al.}{2012}]{Henry2012}
{Henry} A.~L.,  {Martin} C.~L.,  {Dressler} A.,  {Sawicki} M.,   {McCarthy} P.,
   2012, \mn@doi [\apj] {10.1088/0004-637X/744/2/149}, \href
  {http://adsabs.harvard.edu/abs/2012ApJ...744..149H} {744, 149}

\bibitem[\protect\citeauthoryear{{Hu}, {Cowie}, {Capak}, {McMahon}, {Hayashino}
   \& {Komiyama}}{{Hu} et~al.}{2004}]{Hu2004}
{Hu} E.~M.,  {Cowie} L.~L.,  {Capak} P.,  {McMahon} R.~G.,  {Hayashino} T.,
  {Komiyama} Y.,  2004, \mn@doi [\aj] {10.1086/381302}, \href
  {http://adsabs.harvard.edu/abs/2004AJ....127..563H} {127, 563}

\bibitem[\protect\citeauthoryear{{Johnston}}{{Johnston}}{2011}]{Johnston2011}
{Johnston} R.,  2011, \mn@doi [\aapr] {10.1007/s00159-011-0041-9}, \href
  {http://adsabs.harvard.edu/abs/2011A%26ARv..19...41J} {19, 41}

\bibitem[\protect\citeauthoryear{{Konno}, {Ouchi}, {Nakajima}, {Duval},
  {Kusakabe}, {Ono}  \& {Shimasaku}}{{Konno} et~al.}{2016}]{Konno2015}
{Konno} A.,  {Ouchi} M.,  {Nakajima} K.,  {Duval} F.,  {Kusakabe} H.,  {Ono}
  Y.,   {Shimasaku} K.,  2016, \mn@doi [\apj] {10.3847/0004-637X/823/1/20},
  \href {http://adsabs.harvard.edu/abs/2016ApJ...823...20K} {823, 20}

\bibitem[\protect\citeauthoryear{{Kurk}, {Cimatti}, {di Serego Alighieri},
  {Vernet}, {Daddi}, {Ferrara}  \& {Ciardi}}{{Kurk} et~al.}{2004}]{Kurk2004}
{Kurk} J.~D.,  {Cimatti} A.,  {di Serego Alighieri} S.,  {Vernet} J.,  {Daddi}
  E.,  {Ferrara} A.,   {Ciardi} B.,  2004, \mn@doi [\aap]
  {10.1051/0004-6361:20040189}, \href
  {http://adsabs.harvard.edu/abs/2004A%26A...422L..13K} {422, L13}

\bibitem[\protect\citeauthoryear{{Matsuda} et~al.,}{{Matsuda}
  et~al.}{2012}]{Matsuda2012}
{Matsuda} Y.,  et~al., 2012, \mn@doi [\mnras]
  {10.1111/j.1365-2966.2012.21143.x}, \href
  {http://adsabs.harvard.edu/abs/2012MNRAS.425..878M} {425, 878}

\bibitem[\protect\citeauthoryear{{Matthee}, {Sobral}, {Santos},
  {R{\"o}ttgering}, {Darvish}  \& {Mobasher}}{{Matthee}
  et~al.}{2015}]{Matthee2015}
{Matthee} J.,  {Sobral} D.,  {Santos} S.,  {R{\"o}ttgering} H.,  {Darvish} B.,
   {Mobasher} B.,  2015, \mn@doi [\mnras] {10.1093/mnras/stv947}, \href
  {http://adsabs.harvard.edu/abs/2015MNRAS.451..400M} {451, 400}

\bibitem[\protect\citeauthoryear{{Momose} et~al.,}{{Momose}
  et~al.}{2014}]{Momose2014}
{Momose} R.,  et~al., 2014, \mn@doi [\mnras] {10.1093/mnras/stu825}, \href
  {http://adsabs.harvard.edu/abs/2014MNRAS.442..110M} {442, 110}

\bibitem[\protect\citeauthoryear{{Ouchi} et~al.,}{{Ouchi}
  et~al.}{2003}]{Ouchi2003}
{Ouchi} M.,  et~al., 2003, \mn@doi [\apj] {10.1086/344476}, \href
  {http://adsabs.harvard.edu/abs/2003ApJ...582...60O} {582, 60}

\bibitem[\protect\citeauthoryear{{Ouchi} et~al.,}{{Ouchi}
  et~al.}{2008}]{Ouchi2008}
{Ouchi} M.,  et~al., 2008, \mn@doi [\apjs] {10.1086/527673}, \href
  {http://adsabs.harvard.edu/abs/2008ApJS..176..301O} {176, 301}

\bibitem[\protect\citeauthoryear{{Rauch} et~al.,}{{Rauch}
  et~al.}{2008}]{Rauch2008}
{Rauch} M.,  et~al., 2008, \mn@doi [\apj] {10.1086/525846}, \href
  {http://adsabs.harvard.edu/abs/2008ApJ...681..856R} {681, 856}

\bibitem[\protect\citeauthoryear{{Rhoads}, {Malhotra}, {Dey}, {Stern},
  {Spinrad}  \& {Jannuzi}}{{Rhoads} et~al.}{2000}]{Rhoads2000}
{Rhoads} J.~E.,  {Malhotra} S.,  {Dey} A.,  {Stern} D.,  {Spinrad} H.,
  {Jannuzi} B.~T.,  2000, \mn@doi [\apjl] {10.1086/317874}, \href
  {http://adsabs.harvard.edu/abs/2000ApJ...545L..85R} {545, L85}

\bibitem[\protect\citeauthoryear{{Santos}, {Sobral}  \& {Matthee}}{{Santos}
  et~al.}{2016}]{Santos2016}
{Santos} S.,  {Sobral} D.,   {Matthee} J.,  2016, preprint, \href
  {http://adsabs.harvard.edu/abs/2016arXiv160607435S} {} (\mn@eprint {arXiv}
  {1606.07435})

\bibitem[\protect\citeauthoryear{{Schaerer}, {Hayes}, {Verhamme}  \&
  {Teyssier}}{{Schaerer} et~al.}{2011}]{Schaerer2011}
{Schaerer} D.,  {Hayes} M.,  {Verhamme} A.,   {Teyssier} R.,  2011, \mn@doi
  [\aap] {10.1051/0004-6361/201116709}, \href
  {http://adsabs.harvard.edu/abs/2011A%26A...531A..12S} {531, A12}

\bibitem[\protect\citeauthoryear{{Shimasaku} et~al.,}{{Shimasaku}
  et~al.}{2006}]{Shimasaku2006}
{Shimasaku} K.,  et~al., 2006, \mn@doi [\pasj] {10.1093/pasj/58.2.313}, \href
  {http://adsabs.harvard.edu/abs/2006PASJ...58..313S} {58, 313}

\bibitem[\protect\citeauthoryear{{Sobral} et~al.,}{{Sobral}
  et~al.}{2009}]{Sobral2009}
{Sobral} D.,  et~al., 2009, \mn@doi [\mnras]
  {10.1111/j.1365-2966.2009.15129.x}, \href
  {http://adsabs.harvard.edu/abs/2009MNRAS.398...75S} {398, 75}

\bibitem[\protect\citeauthoryear{{Sobral}, {Smail}, {Best}, {Geach}, {Matsuda},
  {Stott}, {Cirasuolo}  \& {Kurk}}{{Sobral} et~al.}{2013}]{Sobral2013}
{Sobral} D.,  {Smail} I.,  {Best} P.~N.,  {Geach} J.~E.,  {Matsuda} Y.,
  {Stott} J.~P.,  {Cirasuolo} M.,   {Kurk} J.,  2013, \mn@doi [\mnras]
  {10.1093/mnras/sts096}, \href
  {http://adsabs.harvard.edu/abs/2013MNRAS.428.1128S} {428, 1128}

\bibitem[\protect\citeauthoryear{{Tremonti} et~al.,}{{Tremonti}
  et~al.}{2004}]{Tremonti2004}
{Tremonti} C.~A.,  et~al., 2004, \mn@doi [\apj] {10.1086/423264}, \href
  {http://adsabs.harvard.edu/abs/2004ApJ...613..898T} {613, 898}

\bibitem[\protect\citeauthoryear{{Verhamme}, {Schaerer}  \&
  {Maselli}}{{Verhamme} et~al.}{2006}]{Verhamme2006}
{Verhamme} A.,  {Schaerer} D.,   {Maselli} A.,  2006, \mn@doi [\aap]
  {10.1051/0004-6361:20065554}, \href
  {http://adsabs.harvard.edu/abs/2006A%26A...460..397V} {460, 397}

\bibitem[\protect\citeauthoryear{{Wisotzki} et~al.,}{{Wisotzki}
  et~al.}{2016}]{Wisotzki2016}
{Wisotzki} L.,  et~al., 2016, \mn@doi [\aap] {10.1051/0004-6361/201527384},
  \href {http://adsabs.harvard.edu/abs/2016A%26A...587A..98W} {587, A98}

\bibitem[\protect\citeauthoryear{{Yamada}, {Matsuda}, {Kousai}, {Hayashino},
  {Morimoto}  \& {Umemura}}{{Yamada} et~al.}{2012}]{Yamada2012}
{Yamada} T.,  {Matsuda} Y.,  {Kousai} K.,  {Hayashino} T.,  {Morimoto} N.,
  {Umemura} M.,  2012, \mn@doi [\apj] {10.1088/0004-637X/751/1/29}, \href
  {http://adsabs.harvard.edu/abs/2012ApJ...751...29Y} {751, 29}

\bibitem[\protect\citeauthoryear{{Yuma} et~al.,}{{Yuma}
  et~al.}{2013}]{Yuma2013}
{Yuma} S.,  et~al., 2013, \mn@doi [\apj] {10.1088/0004-637X/779/1/53}, \href
  {http://adsabs.harvard.edu/abs/2013ApJ...779...53Y} {779, 53}

\bibitem[\protect\citeauthoryear{{van Breukelen}, {Jarvis}  \& {Venemans}}{{van
  Breukelen} et~al.}{2005}]{vanBreukelen2005}
{van Breukelen} C.,  {Jarvis} M.~J.,   {Venemans} B.~P.,  2005, \mn@doi
  [\mnras] {10.1111/j.1365-2966.2005.08916.x}, \href
  {http://adsabs.harvard.edu/abs/2005MNRAS.359..895V} {359, 895}

\makeatother
\end{thebibliography}

\appendix
\section{Flux Catalogue}
\label{blah}

Table A1 presents various flux estimates for all 59 objects
  detected automatically with {\sc{muselet}}. The first column gives
  the ID of the object in B15. The second and third columns
  give the RA and DEC coordinates of each detection as found by
  {\sc{muselet}}, the fourth column gives the peak wavelength
  $\lambda$ of {\sc{muselet}}'s detection, and the fifth column gives
  the Ly$\alpha$ redshift. The following columns give four
  different flux estimates for each source, in column 6 the B15 flux
  measured via {\sc{platefit}}, in column 7 the curve of growth flux
  measured in Wisotzki et al. (2016; where given), column 8
  gives the two arcsecond aperture flux estimate from this work, and
the 9th column gives the curve of growth flux estimate from this
work. All fluxes are quoted in units of
10$^{-18}$ergs$^{-1}$cm$^{-2}$. In the 10th, and final, column, we
give the diameter of the aperture
within which we make the flux measurement in the curve-of-growth
analysis.

\onecolumn
\begin{landscape}

\begin{center}

\begin{longtable}{cccccrrrrc} 
\caption{Various flux estimates for all 59 objects
  detected automatically with {\sc{muselet}}. Column descriptions are
  given at the beginning of the Appendix.}\\
\label{tab:fluxes}

 {\bf{Bacon ID}} & {\bf{RA$_{\rm{{MUSELET}}}$}} &
 {\bf{DEC$_{\rm{{MUSELET}}}$}} & {\bf{$\lambda_{{\rm{obs\;
           MUSELET}}}$}} & {\bf{ z$_{{\rm{Ly \alpha}}}$}} &
 {\bf{Flux$_{{\rm{B15}}}$}} & {\bf{Flux$_{{\rm{Wisotzki}}}$}}   &
 {\bf{Flux$_{{\rm{2\arcsec}}}$}} & {\bf{Flux$_{{\rm{C.o.G}}}$}} &
 {\bf{Aperture$_{{\rm{diameter\,C.o.G\,\arcsec}}}$}} \\

& (J2000) & (J2000) & \AA & & 10$^{-18}$ergs$^{-1}$cm$^{-2}$ & 10$^{-18}$ergs$^{-1}$cm$^{-2}$  & 10$^{-18}$ergs$^{-1}$cm$^{-2}$ & 10$^{-18}$ergs$^{-1}$cm$^{-2}$ &  \\
\hline
\hline

\endhead
 43 & 338.2168 & -60.5618 &  5217.50 &  3.29 &  19.90 &  34.00 &  15.04$\pm$ 2.61 &  29.28$\pm$ 5.88 &  4.04 \\ 
 71 & 338.2234 & -60.5652 &  4967.50 &  3.09 &  23.01 &  -- &  21.87$\pm$ 4.01 &  30.39$\pm$ 7.34 &  5.66 \\ 
 92 & 338.2283 & -60.5706 &  6783.75 &  4.58 &  10.45 &  22.20 &  9.70$\pm$ 2.14 &  16.32$\pm$ 4.29 &  4.04 \\ 
 95 & 338.2441 & -60.5691 &  6350.00 &  4.22 &  3.71 &  12.70 &  3.58$\pm$ 2.51 &  9.03$\pm$ 7.64 &  4.85 \\ 
 112 & 338.2396 & -60.5635 &  5967.50 &  3.91 &  14.43 &  26.40 &  15.65$\pm$ 2.35 &  29.06$\pm$ 11.84 &  7.68 \\ 
 139 & 338.2310 & -60.5611 &  5288.75 &  3.35 &  6.05 &  19.60 &  10.20$\pm$ 2.27 &  26.72$\pm$ 12.83 &  8.08 \\ 
 144 & 338.2455 & -60.5666 &  6098.75 &  4.02 &  31.14 &  -- &  31.29$\pm$ 4.64 &  43.51$\pm$ 6.06 &  4.04 \\ 
 159 & 338.2469 & -60.5630 &  5772.50 &  3.75 &  4.31 &  -- &  5.70$\pm$ 1.61 &  11.24$\pm$ 5.34 &  4.04 \\ 
 162 & 338.2460 & -60.5561 &  5201.25 &  3.28 &  1.29 &  -- &  3.51$\pm$ 1.82 &  4.61$\pm$ 4.64 &  4.04 \\ 
 181 & 338.2462 & -60.5571 &  5273.75 &  3.34 &  18.15 &  27.10 &  18.62$\pm$ 4.12 &  28.42$\pm$ 6.50 &  5.25 \\ 
 202 & 338.2473 & -60.5685 &  5201.25 &  3.28 &  10.94 &  -- &  15.24$\pm$ 2.31 &  19.94$\pm$ 6.07 &  4.04 \\ 
 216 & 338.2363 & -60.5607 &  6100.00 &  4.02 &  6.55 &  12.90 &  9.87$\pm$ 1.46 &  17.10$\pm$ 7.04 &  5.66 \\ 
 218 & 338.2414 & -60.5556 &  7173.75 &  4.90 &  2.69 &  -- &  3.87$\pm$ 1.34 &  5.89$\pm$ 3.97 &  3.64 \\ 
 225 & 338.2318 & -60.5553 &  6458.75 &  4.31 &  8.72 &  -- &  10.42$\pm$ 1.97 &  13.49$\pm$ 5.82 &  4.44 \\ 
 232 & 338.2191 & -60.5610 &  7557.50 &  5.22 &  2.04 &  4.30 &  2.56$\pm$ 1.07 &  3.44$\pm$ 2.45 &  4.04 \\ 
 246 & 338.2350 & -60.5584 &  8121.25 &  5.68 &  5.68 &  12.60 &  7.12$\pm$ 3.35 &  11.85$\pm$ 7.24 &  5.25 \\ 
 290 & 338.2425 & -60.5612 &  8618.75 &  6.09 &  4.08 &  -- &  4.64$\pm$ 2.14 &  6.19$\pm$ 8.27 &  4.85 \\ 
 294 & 338.2196 & -60.5596 &  6070.00 &  3.99 &  3.26 &  7.30 &  4.16$\pm$ 1.14 &  4.87$\pm$ 2.11 &  2.83 \\ 
 308 & 338.2416 & -60.5617 &  6101.25 &  4.02 &  2.26 &  7.60 &  4.95$\pm$ 1.45 &  6.33$\pm$ 3.54 &  3.64 \\ 
 311 & 338.2383 & -60.5643 &  5945.00 &  3.89 &  3.24 &  5.00 &  4.14$\pm$ 1.16 &  4.49$\pm$ 2.55 &  3.23 \\ 
 325 & 338.2179 & -60.5629 &  6931.25 &  4.70 &  6.17 &  11.50 &  8.16$\pm$ 1.48 &  10.89$\pm$ 3.39 &  4.04 \\ 
 334 & 338.2202 & -60.5598 &  7192.50 &  4.91 &  2.77 &  -- &  3.11$\pm$ 0.90 &  3.48$\pm$ 1.36 &  2.83 \\ 
 338 & 338.2171 & -60.5560 &  7173.75 &  4.90 &  5.74 &  -- &  4.04$\pm$ 1.76 &  3.96$\pm$ 1.30 &  1.62 \\ 
 393 & 338.2414 & -60.5581 &  6308.75 &  4.19 &  3.22 &  7.10 &  5.92$\pm$ 2.94 &  14.39$\pm$ 8.94 &  4.04 \\ 
 422 & 338.2173 & -60.5697 &  5021.25 &  3.13 &  2.82 &  6.60 &  5.09$\pm$ 2.09 &  5.86$\pm$ 3.61 &  2.83 \\ 
 430 & 338.2331 & -60.5644 &  8855.00 &  6.28 &  4.77 &  -- &  6.89$\pm$ 2.60 &  10.12$\pm$ 7.51 &  4.04 \\ 
 433 & 338.2152 & -60.5583 &  5435.00 &  3.47 &  7.51 &  -- &  9.58$\pm$ 1.67 &  12.69$\pm$ 5.17 &  4.04 \\ 
 437 & 338.2326 & -60.5686 &  5010.00 &  3.12 &  7.38 &  10.90 &  7.01$\pm$ 2.45 &  10.14$\pm$ 4.46 &  4.04 \\ 
 441 & 338.2279 & -60.5651 &  6923.75 &  4.69 &  6.54 &  -- &  5.78$\pm$ 1.54 &  7.88$\pm$ 3.09 &  3.64 \\ 
 449 & 338.2437 & -60.5672 &  5200.00 &  3.28 &  3.59 &  -- &  4.94$\pm$ 1.53 &  4.82$\pm$ 1.53 &  2.02 \\ 
 453 & 338.2298 & -60.5647 &  6933.75 &  4.70 &  1.46 &  -- &  1.64$\pm$ 0.72 &  1.18$\pm$ 1.24 &  2.83 \\ 
 462 & 338.2392 & -60.5607 &  8021.25 &  5.60 &  3.68 &  -- &  1.37$\pm$ 2.18 &  1.35$\pm$ 2.69 &  2.42 \\ 
 469 & 338.2340 & -60.5643 &  5431.25 &  3.47 &  2.64 &  -- &  2.88$\pm$ 1.20 &  4.97$\pm$ 5.00 &  4.04 \\ 
 478 & 338.2160 & -60.5577 &  5435.00 &  3.47 &  2.41 &  -- &  3.03$\pm$ 1.75 &  7.28$\pm$ 9.27 &  5.66 \\ 
 484 & 338.2441 & -60.5702 &  7190.00 &  4.91 &  2.94 &  -- &  2.90$\pm$ 0.96 &  3.03$\pm$ 1.32 &  2.42 \\ 
 489 & 338.2377 & -60.5623 &  4810.00 &  2.96 &  2.65 &  5.10 &  4.76$\pm$ 1.76 &  5.98$\pm$ 3.20 &  2.83 \\ 
 492 & 338.2411 & -60.5662 &  8221.25 &  5.76 &  2.88 &  -- &  4.06$\pm$ 1.41 &  5.34$\pm$ 3.81 &  4.04 \\ 
 498 & 338.2479 & -60.5693 &  6330.00 &  4.21 &  3.61 &  -- &  4.18$\pm$ 1.76 &  4.73$\pm$ 3.06 &  2.83 \\ 
 499 & 338.2279 & -60.5651 &  6923.75 &  4.69 &  6.40 &  -- &  5.78$\pm$ 1.54 &  7.88$\pm$ 3.09 &  3.64 \\ 
 500 & 338.2364 & -60.5656 &  5475.00 &  3.50 &  1.85 &  -- &  2.45$\pm$ 1.62 &  2.38$\pm$ 3.49 &  3.64 \\ 
 503 & 338.2350 & -60.5640 &  5287.50 &  3.35 &  3.43 &  -- &  2.99$\pm$ 1.59 &  3.52$\pm$ 2.44 &  2.83 \\ 
 513 & 338.2478 & -60.5589 &  5202.50 &  3.28 &  3.09 &  -- &  3.01$\pm$ 2.22 &  5.87$\pm$ 10.62 &  4.85 \\ 
 546 & 338.2238 & -60.5614 &  8162.50 &  5.71 &  3.80 &  8.00 &  6.40$\pm$ 1.35 &  9.06$\pm$ 3.95 &  3.64 \\ 
 547 & 338.2247 & -60.5683 &  8161.25 &  5.71 &  3.56 &  10.70 &  6.52$\pm$ 1.23 &  9.71$\pm$ 4.19 &  4.04 \\ 
 549 & 338.2318 & -60.5613 &  6900.00 &  4.67 &  2.38 &  4.90 &  3.95$\pm$ 1.18 &  4.13$\pm$ 1.70 &  2.42 \\ 
 551 & 338.2296 & -60.5670 &  5083.75 &  3.18 &  4.04 &  -- &  5.13$\pm$ 1.45 &  6.59$\pm$ 6.76 &  5.25 \\ 
 552 & 338.2218 & -60.5630 &  7392.50 &  5.08 &  1.76 &  -- &  2.05$\pm$ 1.67 &  2.01$\pm$ 2.56 &  2.83 \\ 
 553 & 338.2193 & -60.5655 &  7392.50 &  5.08 &  4.69 &  9.30 &  6.53$\pm$ 1.50 &  11.00$\pm$ 7.26 &  6.46 \\ 
 555 & 338.2398 & -60.5651 &  6700.00 &  4.51 &  1.05 &  -- &  2.44$\pm$ 1.24 &  5.64$\pm$ 4.71 &  4.85 \\ 
 557 & 338.2240 & -60.5633 &  7542.50 &  5.20 &  1.90 &  -- &  2.71$\pm$ 1.21 &  2.76$\pm$ 2.62 &  3.23 \\ 
 558 & 338.2267 & -60.5672 &  5018.75 &  3.13 &  2.85 &  6.10 &  3.78$\pm$ 1.22 &  6.20$\pm$ 5.38 &  5.25 \\ 
 560 & 338.2464 & -60.5568 &  8363.75 &  5.88 &  5.38 &  -- &  6.58$\pm$ 1.60 &  8.51$\pm$ 2.29 &  2.83 \\ 
 561 & 338.2155 & -60.5610 &  7065.00 &  4.81 &  1.78 &  -- &  2.50$\pm$ 1.08 &  2.47$\pm$ 1.38 &  2.42 \\ 
 563 & 338.2182 & -60.5669 &  5868.75 &  3.83 &  3.68 &  6.60 &  3.54$\pm$ 1.74 &  4.25$\pm$ 5.94 &  4.44 \\ 
 568 & 338.2240 & -60.5598 &  6886.25 &  4.66 &  3.42 &  4.60 &  4.63$\pm$ 1.29 &  6.03$\pm$ 4.88 &  4.44 \\ 
 573 & 338.2478 & -60.5703 &  8842.50 &  6.27 &  2.64 &  -- &  4.68$\pm$ 3.80 &  4.67$\pm$ 3.16 &  1.62 \\ 
 577 & 338.2419 & -60.5670 &  8221.25 &  5.76 &  6.55 &  -- &  9.68$\pm$ 1.35 &  13.90$\pm$ 3.62 &  4.04 \\ 
 578 & 338.2296 & -60.5670 &  5083.75 &  3.18 &  2.75 &  -- &  5.13$\pm$ 1.45 &  6.59$\pm$ 6.76 &  5.25 \\ 
 585 & 338.2389 & -60.5631 &  5275.00 &  3.34 &  2.55 &  -- &  2.19$\pm$ 1.26 &  2.81$\pm$ 1.97 &  2.83 \\ 
& &  &&&&&& &\\ 
\end{longtable}
\end{center}
\end{landscape}

\twocolumn

\label{lastpage}

\end{document}